\documentclass[12pt]{article}
\pdfoutput=1

\usepackage{jcappub}

\usepackage{amsmath,amsfonts}
\usepackage{amssymb}
\usepackage[off]{auto-pst-pdf}
\usepackage{graphics}
\usepackage{psfrag}
\usepackage{multirow}

\usepackage{fullpage}

\makeatletter
\@addtoreset{equation}{section}
\makeatother

\newcommand{\be}{\begin{equation}}
\newcommand{\ee}{\end{equation}}
\newcommand{\bea}{\begin{eqnarray}}
\newcommand{\eea}{\end{eqnarray}}

\newcommand{\tr}{\operatorname{tr}}

\newcommand{\nn}{\nonumber}

\renewcommand{\sl}{/\!\!\!}
\newcommand{\Dsl}{\,\sl\!D}
\newcommand{\Q}{\mathcal Q}
\newcommand{\A}{\mathcal A}
\renewcommand{\S}{\mathcal S}
\renewcommand{\H}{\mathcal H}
\newcommand{\F}{\mathcal F}
\renewcommand{\t}{\bf t}

\newcommand{\bmat}{\begin{pmatrix}}
\newcommand{\emat}{\end{pmatrix}}

\newcommand{\mumin}{\tilde \mu_{\rm eff}}
\newcommand{\muext}{\mu_{\rm eff}}

\begin{document}

\title{The Dynamical Composite Higgs}

\author{Gero von Gersdorff,}
\emailAdd{gersdorff@gmail.com}

\author{Eduardo Pont\'on and}
\emailAdd{eponton@ift.unesp.br}

\author{Rogerio Rosenfeld}
\emailAdd{rosenfel@ift.unesp.br}

\affiliation{ICTP South American Institute for Fundamental
  Research \& Instituto de F\'isica Te\'orica \\ 
  Universidade Estadual  Paulista, Rua Dr. Bento T. Ferraz, S\~ao Paulo, Brazil}

\abstract{
We present a simple microscopic realization of a pseudo-Nambu-Goldstone
(pNGB) boson Higgs scenario arising from the breaking of $SO(5) \to
SO(4)$. The Higgs constituents are explicitly identified as well as the
interactions responsible for forming the bound state and breaking the
electroweak symmetry. This outcome follows from the presence of
four-fermion interactions with a super-critical coupling, and uses the
Nambu-Jona-Lasinio mechanism to break the global $SO(5)$ symmetry. The
Higgs potential is found to be insensitive to high energy scales due to
the existence of an approximate infrared fixed point.
The appearance of vector resonances is described and the
correspondence with other proposals in the literature established. 
The model described here is significantly simpler than
other recent ultraviolet completions of pNGB scenarios.
The observed Higgs mass can be accommodated, and agreement with
electroweak precision tests achieved in certain regions of parameter
space.
There are also new vector-like fermions, some of which may lie within reach of the LHC.
In addition, we predict a heavy standard model singlet scalar in the multi-TeV range.
The amount of fine-tuning required in the model is studied.
Finally, we show that such a scheme can be completed in the ultraviolet
by a renormalizable theory.

}

{\center \today}

\maketitle

\section{Introduction}

Theories with fundamental scalars usually suffer from a quadratic sensitivity to the ultraviolet (UV)
that leads to naturalness problems when hierarchically separated energy scales
are present. Historically this has motivated extensions
of the Standard Model with either new symmetries that protect the theory
against short distance details ({\it e.g.} supersymmetric models) or with an UV completion where the scalar 
particle is not fundamental ({\it e.g.}  technicolor (TC) models).

The usual TC models were constructed in analogy with QCD and the Higgs was identified with the elusive $\sigma$-meson.
This interpretation is now excluded since the Higgs was found to be light and narrow~\cite{Aad:2012tfa,Chatrchyan:2012ufa}.  However, it was also proposed in the 1980's that the Higgs boson could be a pseudo-Nambu-Goldstone (pNGB) composite 
state from the spontaneous breaking of a global symmetry~\cite{Kaplan:1983sm,Georgi:1984af,Georgi:1984ef,Dugan:1984hq} (for a nice review, see~\cite{Georgi:2007zza}).
The study of effective theories of this sort eventually led to what is generically known as composite Higgs (CH) models. 
CH models experienced a revival about a decade ago with their incorporation into extra-dimensional theories in AdS spaces~\cite{Contino:2003ve,Agashe:2004rs}.

Several CH models have been proposed recently and it is timely to explore what are the 
requirements on the underlying theory that gives rise to these effective theories at low energies, generating
the correct electroweak symmetry breaking (EWSB) scale and the Higgs mass. Here we start exploring this question by describing a concrete non-supersymmetric UV completion with the required low-energy properties (see also~\cite{Cacciapaglia:2014uja,Lane:2014vca}). A number of supersymmetric UV completions of CH scenarios have also been proposed~\cite{Kitano:2012wv,Caracciolo:2012je,Parolini:2014rza}.

Implicit in much of the literature is the idea that the underlying physics involves strong dynamics, which eventually triggers
symmetry breaking and the formation of bound states. One can then try to model these strong interactions
as effective 4-fermion interactions and study them in the context of
the Nambu-Jona-Lasinio (NJL) framework.
In the incarnation studied in this paper one has three energy scales: a scale $\Lambda$ associated with
the mass of heavy degrees of freedom that, when integrated out, generate the 4-fermion interactions with 
strength $G = {\cal O}(1/\Lambda^2)$ --we will sometimes refer to it as the compositeness scale;
a scale $f$ associated with the breaking of a global symmetry through the vacuum expectation value (vev) of an effective
scalar field that generates Goldstone bosons; and finally $v$, the scale of electroweak (EW) breaking that is 
generated through a 1-loop effective potential from (small) effects that violate 
the global symmetry explicitly.

We are largely inspired by the seminal paper by Bardeen, Hill and Lindner~\cite{Bardeen:1989ds} 
(see also the early works~\cite{Terazawa:1976xx,Nambu:1989jt,Miransky:1988xi,Miransky:1989ds})
who studied a model where the Higgs boson is a composite state of top quarks, arising from 4-fermion top-quark interactions \`a la NJL~\cite{Nambu:1961tp,Nambu:1961fr}.  
There, a self-consistent solution of the gap equation with a nonzero
$\langle \bar{t}t \rangle$ condensate  for a strong enough coupling 
breaks a global symmetry $SU(2)_L\times SU(2)_R \rightarrow SU(2)_V$
generating  the Goldstone bosons that are absorbed by the EW gauge bosons
in the usual symmetry breaking mechanism, which in this case has an explicit custodial symmetry.
The solution to the gap equation determines the mass
of the top quark and predicts a new composite scalar particle,
associated with the bound state of $\bar{t}t$ that is
identified with the Higgs boson.
In the approximation where gauge loops are neglected, the Higgs mass is predicted to be $m_h \approx 1.32\, m_t$~\cite{Bardeen:1989ds}. 
Moreover, at the scale $\Lambda$ the scalar kinetic term approaches zero (the ``compositeness condition''), 
and hence the top Yukawa coupling reaches a Landau pole. 
It turns out that this is only possible for a rather heavy top quark.
In spite of its attractiveness as a model of dynamical symmetry breaking
without fundamental scalars, the most minimal model fails since it gives both a
too heavy top quarks and a too heavy Higgs boson.

Prior to the Higgs discovery, the problem with the top quark mass was solved in models
that include the  top quark see-saw (TSS) mechanism~\cite{Dobrescu:1997nm,Chivukula:1998wd}. However, after the
2012 discovery we know that the Higgs boson is lighter than the top quark. As a result, the problem of too heavy a Higgs in NJL-type models (with a TSS) has been brought up again recently.
In particular, the idea to enlarge the pattern of global symmetry breaking in order 
to realize the Higgs boson as a pNGB has been
put forward in~\cite{Cheng:2013qwa,Barnard:2013zea,Fukano:2013aea,Fukano:2014zpa,Cheng:2014dwa,Fukano:2014dta}.

In this paper we consider a UV complete scenario that realizes the Higgs as a pNGB of the symmetry breaking $SO(5) \to SO(4)$. As is well-known~\cite{Agashe:2004rs} , this leads to a minimal set of NGB's while containing a custodial symmetry~\cite{Agashe:2003zs} and allowing for the ``custodial protection'' of the $Zb_L\bar{b}_L$ vertex~\cite{Agashe:2006at}. 
We point out that four-fermion interactions, which may easily arise from a renormalizable model, can naturally realize the above symmetry breaking pattern. This leads to simpler realizations than have been explored in the literature, and goes beyond the CCWZ~\cite{Coleman:1969sm,Callan:1969sn} non-linear realizations  of the symmetry employed in most of the composite Higgs literature. We also show that the heavy spin-1 sector, which would be responsible for cancelling the quadratic divergences in the Higgs mass parameter due to the SM gauge bosons, arises quite naturally in this framework. In regards to fermions, we focus on the top sector and proceed guided by the principle of minimality. We find  that the most minimal model is typically ruled out by electroweak precision tests (EWPT)  due to a negative 1-loop contribution to the Peskin-Takeuchi $T$-parameter~\cite{Peskin:1991sw}. However, a slightly extended version allows agreement with precision measurements. We find that it is necessary to include a moderate amount of (soft) custodial breaking in the BSM sector which, however, does not spoil the calculability implied by the custodial $SO(4)$. We also compute the dynamically generated Higgs potential, and explore the region of parameter space where the observed Higgs mass can be reproduced. It is interesting that the presence of IR quasi-fixed points make the low-energy predictions largely insensitive to the uncertainties associated with the underlying strong dynamics.

This paper is organized as follows. In Section~\ref{Model}, we review the NJL symmetry breaking mechanism as applied to the present $SO(5)$ scenario. We remind the reader of the IR quasi-fixed point that relates the fermion and (radial) scalar masses. We also present the minimal fermionic sector that leads to a realistic low-energy field content, as well as other natural extensions inspired by an analogy with models of partial compositeness (on which we also comment). Spin-1 resonances are then introduced and shown to be naturally  associated with a ``hidden local symmetry''.

Sections~\ref{GBpot} and \ref{sec:ewsb} are devoted to the dynamical breaking of the EW symmetry. This arises from the explicit breaking of the $SO(5)$ symmetry by the gauge interactions, as well as by certain mass terms in the fermionic sector. Of particular note is  that the $SO(5)$ breaking is not only soft (so that the quadratic sensitivity to the UV of the weak scale is absent), but that there is a second IR quasi-fixed point that largely shields the low-energy predictions from the UV details. We also study the interplay between EWSB and EWPT. In Section~\ref{naturalness} we analyze the resulting tuning in our scenario, which itself depends on a relatively small number of parameters. Interestingly, we find that although the new resonances are typically in the TeV range, the theory could be technically natural due to a nearby enhanced symmetry point. Finally, in Section~\ref{SO(N)}, we present a simple renormalizable model that  can serve as a UV completion to our present work, generating the necessary four-fermion operators with the required $SO(5)$ symmetry and in Section~\ref{pheno} we present a few comments on the expectations for present and future colliders within our scenario.   Section~\ref{conclusions} contains our summary and conclusions.

We also include a number of appendices. In App.~\ref{app:SO(N)} we present details of the Fierz rearrangements necessary to relate the renormalizable theory of Section~\ref{SO(N)} and the $SO(N)$ NJL model. We summarize our $SO(5)$ conventions in App.~\ref{GroupTheory}, while in App.~\ref{app:potential} we provide a few technical details of the evaluation of the Higgs potential due to the spin-0, spin-1/2 and spin-1 resonances. Finally, App.~\ref{app:IRRegulation} discusses a technical point related to the IR divergences that appear when expanding the potential in powers of the Higgs vev, and App.~\ref{app:RGeqs} summarizes the RG equations that are used to establish the presence of the IR quasi-fixed points mentioned above.

\section{Description of the Model}
\label{Model}

\subsection{NJL Breaking of $SO(5)_L \times U(1)_X \to SO(4)_L \times U(1)_X$}
\label{NJL}

The breaking of the global group $G = SO(5)_L \times U(1)_X$ can be
achieved by strongly coupled four-fermion operators along the lines of
\cite{Nambu:1961tp,Nambu:1961fr}.  The minimal fermionic field content
to achieve this breaking is comprised of a left handed {$\bf
5_{\frac{2}{3}}$}, denoted by $F_L$, and a right-handed $\bf
1_{\frac{2}{3}}$, denoted by $S_R$.\footnote{\label{SMIssues}The normalization of the
$U(1)_X$ generators is chosen for later convenience, when we discuss
the embedding of the SM electroweak symmetry $SU(2)_L \times U(1)_Y
\subset SO(5)_L \times U(1)_X$, and provide further details of the
fermionic sector necessary to reproduce the observed low-energy
physics.  In addition,  we allow each of the fermion fields to carry
a color index, so that the symmetry of the theory will always contain an additional  $SU(N_c)$ factor. This factor is understood to be weakly gauged and, with $N_c = 3$, will be identified with the QCD interactions. We will assume that all fields are in the fundamental or singlet representations of $SU(N_c)$, depending on whether they are associated with the quarks or leptons of the SM. Except when there could be some ambiguity in how the color indices are contracted, we will not display them explicitly.} The kinetic terms read thus
\be
\mathcal L_{F} ~=~ \sum^5_{j=1} i\bar F^j_L\sl\partial F^j_L+i\bar S_R\sl \partial S_R~.
\label{Lkin}
\ee
This Lagrangian is actually accidentally invariant under the larger
group $G_0\equiv SU(5)_L\times U(1)_L\times U(1)_R$, with $SO(5)_L
\subset SU(5)_L$ and $U(1)_X = [U(1)_L\times U(1)_R]_{\rm diagonal}$,
under which $F_L$ transforms as a $\bf 5_{\frac{2}{3},0}$ and $S_R$ as
a $\bf 1_{0,\frac{2}{3}}$.  However, while the fermionic fields
decompose under $G_0\to G$ simply as
\be
\bf 5_{\bf \frac{2}{3},\bf 0}\to \bf 5_{\frac{2}{3}}\,,\qquad 
\bf 1_{0,\frac{2}{3}}\to \bf 1_{\frac{2}{3}}\,,
\ee 
the decomposition of the composite scalar field $\bar S_R F_L$ is
reducible
\be
\bf 5_{\bf \frac{2}{3},-\bf \frac{2}{3}}\to \bf 5_0+\bf 5_0~,
\label{real}
\ee
where the real and imaginary parts of $\bar S_RF_L$ form two different
irreducible representations of $G$.
As a consequence it is possible to write down two four-fermion
operators that are separately invariant under $G$.  They read
\be
\mathcal L_S=\frac{G_S}{2}\left(\sum_{a=1}^{N_c} \bar S_{R,a} F_L^{i,a}+\bar F_{L,a}^iS^a_R\right)^2\,,\qquad 
\mathcal L_S'=-\frac{G_S'}{2}\left(\sum_{a=1}^{N_c} \bar S_{R,a} F_L^{i,a}-\bar F_{L,a}^iS^a_R\right)^2~,
\label{L4f}
\ee
where, to avoid ambiguity, we also show the color index contractions (see Footnote~\ref{SMIssues}), labeled by the index $a$. Such contractions will be implicitly understood from here on.
For $G_S=G_S'$ the Lagrangian $\mathcal L_S+\mathcal L_S'= 2G_S|\bar
S_RF_L^i|^2$ becomes indeed invariant under the extended group $G_0$,
and hence $G_S\neq G_S'$ parametrizes the explicit breaking $G_0\to
G$.

%
%
%
Whenever the four-fermion couplings $G_S$ or $G_S'$ are greater than some critical value \cite{Nambu:1961fr}, the composites $\bar S_RF^i_L$ become dynamical scalar fields that condense and break the global symmetry.
%
%
For both $G_S\approx G_S'$ super-critical, this would lead to
additional light scalar composites (in particular, a second Higgs doublet
under $SU(2)_L$).  Here we effectively restrict ourselves to the minimal
Lagrangian $\mathcal L_S$ by imposing that only the coupling $G_S$ be
super-critical, while $G_S'$ is assumed to be sub-critical and does not
give rise to any light composites. In other words, the four-fermion interactions of ${\cal L}_S$ lead to a spontaneous symmetry breakdown (reviewed below) that produces light (p)NGB states, while  the interactions in ${\cal L}'_S$ do not induce any such breaking. If $G'_S$ is not close to criticality, the additional loosely bound fermion bilinears will have masses close to the cutoff, and it may not even be appropriate to think of them as well-defined scalar resonances in their own right. Thus, the assumption that $G_S$ is slightly larger than $G'_S$, and close to criticality, allows us to focus from the
start on a minimal set of light degrees of freedom that are relevant to the
low-energy physics.  We will present in Section~\ref{SO(N)} a simple UV model that realizes this picture.
 Alternatively, one could allow for the additional
bound states from $\mathcal L_S'$ (or even from further four-fermion operators) and then add appropriate symmetry
breaking terms to the low-energy theory to make them sufficiently
heavy.  The latter is essentially the approach taken in
Refs.~\cite{Cheng:2013qwa,Cheng:2014dwa}.  Note that in this second
approach, in general, one would need to take into account such
intermediate thresholds to establish a connection between the
low-energy theory and the theory at the UV scale $\Lambda$, where the
physics is most appropriately described by the four-fermion
interactions discussed above.  By assuming that $G_S'$ is sub-critical,
so that there are no such additional scalar states below $\Lambda$, a
more straightforward connection between the UV and IR can in principle
be established. 

We analyse now the theory described above following closely the methods described in~\cite{Bardeen:1989ds}. This will serve not only as a review, but will also allow us to point out the crucial features that result when applied to our scenario.\footnote{One possibility is to perform an analysis based on the gap equation. Here we follow the alternate approach based on the introduction of an effective scalar field $\Phi$. The two approaches are equivalent in the large $N$ limit, but the latter allows to more easily introduce certain subleading $1/N$ effects, as well as those from the (weak) SM gauging. It also gives rise more directly to a rather transparent physical picture. See~\cite{Bardeen:1989ds} for further details.} The first step is to rewrite the four-fermion Lagrangian in terms of real scalar
auxiliary fields $\Phi^i$ in the $\bf 5_0$ representation of $G$ as
\bea
\mathcal L_S &=& -\frac{1}{2G_S}\Phi^2-\Phi(\bar S_RF_L+{\rm h.c.})~,
\label{aux1}
\eea
which can be seen to be equivalent to the form given earlier after
integrating out the auxiliary scalar fields.  The Lagrangian $\mathcal
L_F+\mathcal L_S$ generates, through fermion loops, a kinetic term for
$\Phi$, thus making this field dynamical at scales sufficiently below
the matching scale $\Lambda$ (at which Eq.~(\ref{aux1})
holds).  They also generate tachyonic corrections to the tree level
$\Phi$ mass, eventually leading to the spontaneous breaking of
$SO(5)_L$ to its subgroup $SO(4)_L$, which we identify with the
$SU(2)_L \times SU(2)_R$ symmetry group of the SM (in the limit that
the hypercharge and Yukawa couplings are turned off).  The Goldstone
bosons of this breaking transform in a $\bf 4$ of $SO(4)_L$ and hence
can be identified with the SM Higgs field.  We will trade the scalar
mass parameter for the physical symmetry breaking scale $\hat
f^2=\langle \Phi \rangle^2$.  Finally, fermion loops also generate a
quartic self coupling, so that at low energies $\mathcal L_S$ reads
\bea
\tilde{\mathcal L}_S &=& \frac{1}{2}(\partial_\mu\Phi)^2 - \frac{1}{4} \lambda \left( \Phi^2 - \hat{f}^2 \right)^2 - \xi \, \Phi\, (\bar S_R F_L+{\rm h.c.})~,
\label{LSHLScanonical}
\eea
where the Yukawa coupling $\xi$ appears after canonical normalization
(we do not change the name for $\Phi$ for notational simplicity).  The
fact that the kinetic term for $\Phi$ vanishes at $\Lambda$ is
equivalent to saying that $\xi$ reaches a Landau pole at that scale, a
condition also known as the compositeness boundary condition.\footnote{For completeness, we remind the reader that $\Phi$ becomes tachyonic provided $G_S \Lambda^2$ is larger than a certain critical value. Here we trade $\Lambda$ (where $\xi$ diverges) in favor of the low-energy value of $\xi$, while $G_S$ is traded for the symmetry breaking scale $f$. As a result, we do not need to know $G_S$ except for the assumption that it should be above criticality.}

\begin{figure}
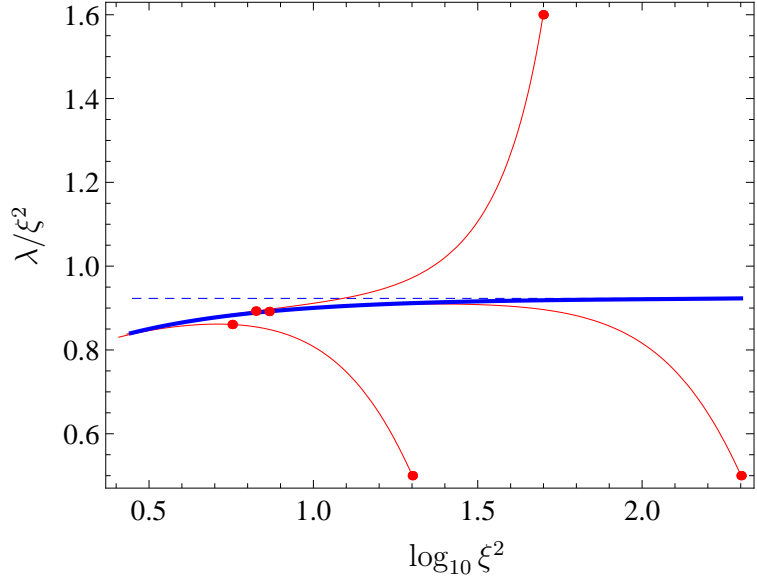

\centering
\psfrag{xxx}{$\log_{10} \xi^2 $}
\psfrag{yyy}{$\lambda/\xi^2$}
\psfragfig[width=0.6\linewidth]{Figures/RGplot}
\caption{RG flow of the couplings $\lambda$ and $\xi$.  The dashed
line is the exact IR fixed point $a_*=12/13$ that is reached in the
absence of gauge interactions.  QCD corrections introduce a mild
dependence $a_*(\xi)$ represented by the solid blue line.  The thin
red lines are examples of trajectories, with the distance between the
dots corresponding to one e-fold of running. For simplicity, we neglect the running of the strong coupling constant.}
\label{RGplot}
\end{figure}

In the NJL model, the quartic self coupling $\lambda$ is a prediction
due to an IR quasi-fixed point for the quantity $\lambda/\xi^2$
\cite{Bardeen:1989ds}.  Indeed, in the $SO(N)$ NJL model, we find
\bea
16 \pi^2\beta_{\xi^2}&=&(4 N_c+N+5)\xi^4~,\nn\\
16 \pi^2\beta_\lambda&=&2(N+8)\lambda^2-8N_c\xi^4 +8N_c\xi^2\lambda~,
\label{betas1}
\eea
which possess an IR stable fixed point at 
\be
\lambda~=~a_* \xi^2~,
\label{FP1}
\ee
where (for $N_c=3$ and $N=5$) $a_*=\frac{12}{13}\approx 0.92$.\footnote{There is also a UV fixed point at $a_* = -1$. In order to reach the IR
fixed point one must have $\lambda/\xi^2 > -1$ at the scale $\Lambda$.}  While
this type of fixed point is present in any scalar-fermion system (even
in the SM~\cite{Pendleton:1980as}), the compositeness condition \--- i.e.~the fact that $\xi$
is strong in the UV \--- guarantees that this fixed point is reached
rapidly in the IR. 
A mild correction to this fixed point behavior is induced by nonzero
gauge couplings.  We illustrate the effect of QCD corrections in
Fig.~\ref{RGplot} (see App.~\ref{app:RGeqs} for the RG equations that include the QCD effects).  
Remarkably, the theory remains completely
predictive even in the presence of gauge corrections, in the sense
that the IR value for $\lambda$ is still fully determined from the IR
value of $\xi$ (see the solid blue line in the figure).  Notice that
the QCD corrections are numerically small, for instance
$a_*(\xi=2)\approx 0.86$, compared to the fixed-point value of
Eqs.~(\ref{betas1}), $a_*\approx 0.92$.

The $SO(5)$ basis employed so far simplifies the description of the
global symmetry breaking, but the SM quantum numbers of the fields are
not manifest.  For the remainder of this paper we thus switch to a
different basis defined as
\be
\left(Q_L^1,\, Q_L^2\right)\equiv \frac{1}{\sqrt{2}}\begin{pmatrix}F_L^4+iF_L^3&F_L^2+iF_L^1\\-F_L^2+iF_L^1&F^4_L-iF_L^3\end{pmatrix}\,,\qquad S_L\equiv F_L^5\,.
\ee
where $Q_L^1$ and $Q_L^2$ transform under $SU(2)_L\times U(1)_Y \subset
SO(5)_L \times U(1)_X$ as $\bf 2_\frac{1}{6}$ and $\bf 2_\frac{7}{6}$
respectively, and $S_L$ (as well as $S_R$) as $\bf 1_{\frac{2}{3}}$.\footnote{See
App.~\ref{GroupTheory}, where we summarize the conventions for the two
different $SO(5)$ bases, as well as the embedding of $SU(2)_L\times U(1)_Y \subset
SO(5)_L \times U(1)_X$.}
Thus, the first doublet $Q_L^1$ has the same quantum numbers of the left-handed
top-bottom doublet, while the second doublet $Q_L^2$ has the exotic
hypercharge $\frac{7}{6}$.  The vector-like singlet $(S_L,S_R)$ has
the same quantum numbers as the right handed top.  In order to obtain
a chiral spectrum which at low energies contains just a left handed $\bf
2_{\frac{1}{6}}$ (to be identified with the SM $(t_L, b_L)$ doublet) and a right-handed $\bf 1_{\frac{2}{3}}$ (to be identified as the SM $t_R$) we will have
to introduce more states in incomplete $G$ multiplets.  We will
describe the top sector in detail in Section~\ref{top}, and continue
focusing in this section on the minimal content required to achieve
the dynamical breaking above.  In the new basis, the scalar sector becomes
\be
\left(\tilde\phi,\, \phi\right)\equiv \frac{1}{\sqrt{2}}\begin{pmatrix}\Phi^4+i\Phi_3&\Phi^2+i\Phi^1\\-\Phi^2+i\Phi^1&\Phi_4-i\Phi^3\end{pmatrix}\,,
\qquad \phi^5\equiv \Phi^5
\ee
where $\phi$ transforms in $\bf 2_\frac{1}{2}$ and $\tilde \phi$ as
$\bf 2_{-\frac{1}{2}}$ of $SU(2)_L\times U(1)_Y$.  The reality property $\Phi^*=\Phi$ translates
into the well-known relation $\phi^*=-i\sigma_2\tilde \phi$.  In this
basis, the Yukawa Lagrangian in Eq.~(\ref{aux1}) reads
\bea
\mathcal L_S &=& -\frac{1}{G_S}\phi^\dagger \phi-\frac{1}{2G_S}\phi_5^2 -(\bar Q^1_LS_R \tilde \phi+\bar Q^2_L S_R\phi+h.c.)-\bar SS\phi_5~,
\label{aux}
\eea
%
while at sufficiently low-energies, where a sizable kinetic term for
$\Phi$ has been radiatively generated, together with quartic scalar
self-interactions, the Lagrangian in Eq.~(\ref{LSHLScanonical}) reads
\bea
\tilde{\mathcal L}_S &=& (\partial_\mu\phi)^2 + \frac{1}{2}(\partial_\mu\phi_5)^2 - \frac{1}{4} \lambda \left( 2\phi^2 + \phi_5^2 - \hat{f}^2 \right)^2 
\nonumber \\
& & \mbox{} - \xi \, \left[(\bar Q^1_LS_R \tilde \phi+\bar Q^2_L S_R\phi+h.c.)-\bar SS\phi_5 \right]~.
\label{LSHLScanonicalNewBasis}
\eea
When $\langle \phi_5 \rangle = \hat{f}$ (and  $\langle \phi \rangle = 0$) , the breaking $SO(5)_L \to
SO(4)_L$ ensues, generating four NGB's that transform as a doublet of
$SU(2)_L$ and can thus be identified with the SM Higgs field.  These
$G/H$ Goldstone bosons are parameterized by introducing a unitary
matrix $U$ as
\be
\hat{\Phi} ~=~ \mathcal H U e_5~,
\label{NGBExp}
\ee
where $\hat{\Phi}=(\tilde\phi,\phi,\phi_5)^T$ and $e_5$ is the unit vector
pointing along the $\phi_5$ direction.  

We have introduced here the
real scalar SM singlet $\H$ that we will refer to as the
\textit{radial mode} in the following, and that acquires the vacuum
expectation value $\langle \H\rangle=\hat f$.  
At the IR quasi-fixed point one has
\be
m^2_\H~=~2\lambda \hat f^2~=~2a_*\xi^2\hat f^2\,,
\label{massH}
\ee
and the mass of the radial mode becomes a prediction of the model once the dynamical fermion mass $m_f
= \xi \hat{f}$ is fixed.  The situation is of course completely
analogous to the relation between the Higgs and top masses in the
seminal paper \cite{Bardeen:1989ds}, where the Higgs boson was identified with the radial mode.

Radiative effects
arising from small explicit $SO(5)_L$ symmetry breaking terms can lead
to a vacuum that is slightly misaligned with the vacuum above, in
which case the EW symmetry will be spontaneously broken.  Such a
misalignment can be parametrized by an angle $s_v \ll 1$ between
$\langle \hat{\Phi} \rangle$ and $\hat{f} e_5 = (0,0,0,0,\hat{f})$, so that
there is a separation of scales between the EW scale and $\hat{f}$
(further details will be discussed later). In this case
\be
\langle \tilde \phi \rangle ~=~ \frac{1}{\sqrt{2}} \begin{pmatrix} s_v \\0 \end{pmatrix} \hat f~, \qquad
\langle \phi \rangle ~=~ \frac{1}{\sqrt{2}} \begin{pmatrix}0\\ s_v\end{pmatrix}\hat f~,\qquad
\langle \phi_5 \rangle ~=~c_v\hat f~.
\ee

The particle content introduced so far allows for precisely one
fermion mass term that is invariant under the SM gauge symmetries,
\be
\mathcal L_F^{\rm mass,0}~=~-\mu_{SS}\bar SS\,.
\label{singletmass}
\ee
It preserves only the subgroup $SO(4)_L$ and hence violates explicitly
the global $SO(5)_L$ symmetry.  It is the analogue of current quark
masses in models of chiral symmetry breaking.  Such a term is in fact
easily seen to be equivalent to a tadpole term for the scalar field
$\phi_5$, via the field redefinition $\phi_5 \to
\phi_5 - \mu_{SS}$ in Eq.~(\ref{aux}) together with (\ref{singletmass}), which eliminates the mass term and instead
generates the tadpole term  (see also Ref.~\cite{Cheng:2013qwa})
\bea
\mathcal L_T &=& \frac{\mu_{SS}}{G_S} \, \phi_5~.
\label{tadpole}
\eea
We define $\tau\equiv\xi\mu_{SS}/{G_S}$, which is the relevant tadpole
term after canonical normalization as in Eq.~(\ref{LSHLScanonical}).
We will see that although EWSB can be achieved without such a term,
the measured value for the Higgs mass can typically only be reproduced
for $\mu_{SS}\neq 0$.  We note that for fixed symmetry breaking scale
$\hat f$, the mass of the radial mode changes to
\be
m^2_\H~=~2 a_*\xi^2\hat f^2+\frac{\tau}{\hat f}~.
\label{massH2}
\ee 
We will see later that this correction is rather small, and the mass
is well approximated by the leading term, Eq.~(\ref{massH}).

\subsection{The Top Sector and Partial Compositeness}
\label{top}

We now complete the model to include a realistic top sector by means
of the TSS mechanism~\cite{Dobrescu:1997nm}.
This bears some resemblance to the partially composite top models
typically introduced in the recent CH literature~\cite{Golterman:2015zwa,Fonseca:2015gva,Barducci:2015oza,Matsedonskyi:2014iha,Barnard:2014tla,Azatov:2014lha,Archer:2014qga,DeCurtis:2014oza,Carena:2014ria};  a more complete set of references to the rather extensive literature can be found in the recent review~\cite{Bellazzini:2014yua}.
We will explicitly make the connection between the two approaches, but
one should keep in mind that in our case all fermions are elementary,
and the distinction between ``composite'' and ``elementary'' is purely
formal.

The fermionic sector of the model so far contains a chiral doublet
with the quantum numbers of the left-handed top-bottom, an exotic (chiral)
left-handed doublet of hypercharge $\frac{7}{6}$, and a
vector-like right handed top.  It is manifestly symmetric under the
global group $G$ [and in particular contains an unbroken custodial
$SO(4)_L$], but does not reduce to the SM top sector at low energies.
The minimal solution is to introduce the right handed fields $Q_R^2$
and $t_R$ with EW quantum numbers $\bf 2_{\frac{7}{6}}$ and $\bf
1_\frac{2}{3}$ respectively, and write the soft $G$-breaking mass
mixing terms:
\be
\mathcal L_F^{mass,1}~=~-\mu'_{QQ}\bar Q_L^2Q_R^2-\mu_{tS}\bar S_Lt_R+{\rm h.c.}
\label{mass1}
\ee
In particular, no new dimensionless $G$-breaking couplings are
introduced.\footnote{Unlike the mass terms in Eq.~(\ref{singletmass})
the operators in Eq.~(\ref{mass1}) are not equivalent to tadpole
terms, as $t_R$ and $Q_R^2$ are not part of the condensate.} One notices that our model thus naturally realizes the TSS
mechanism, in which the top Yukawa coupling arises only after
integrating out the heavy vector-like top $S$~\cite{Dobrescu:1997nm}.

To leading order in $s_v^2$ -- the misalignment angle that parameterizes
EWSB -- the approximate mass eigenvalues in the fermionic sector of this ``minimal" model are
given by
\be
m_S^2~=~\xi^2\hat f^2+\mu_{tS}^2\,,\qquad 
m_Q'^2~=~\mu_{QQ}'^2\,,\qquad 
m_t^2~=~\frac{s_v^2}{2}\frac{\xi^2\hat f^2\,\mu_{tS}^2}{m_S^2}~.
\label{massesf}
\ee
Subleading corrections will in particular split the charge
$\frac{2}{3}$ from the charge $\frac{5}{3}$ states in $Q^2$.
These eigenvalues are not changed by the tadpole term, which only
redefines $\hat{f}$.

A (formally) very similar concept to the TSS mechanism is partial
compositeness (PC) of the top quark (see \cite{Kaplan:1991dc} for an early model and \cite{Contino:2006nn} for a modern view).  
One
distinguishes a $G$-symmetric (or at least $H$-symmetric) sector, made
up of ``composite'' fermions, coupled via mass mixing terms to a set
of ``elementary'' fermions that do not come in complete $G$
multiplets.  The latter, in particular, do not possess direct Yukawa
couplings to the composite Higgs.  In this last sense, one could formally refer to the
fields $Q_L^{1}$, $Q_L^{2}$, $S_L$ and $S_R$ in our setup as ``composite'', and the fields
$Q_R^2$, $t_R$ as ``elementary''.  However, the connection to
other CH constructions based on partial compositeness still would seem to be imperfect, since in such CH models
the elementary sector typically consists of just the SM fermions, and (consequently) the
composite sector is entirely vectorlike. The connection of our setup to other CH models recently considered in the literature can be made more direct by adding the 
$\bf 2_{\frac{1}{6}}$ states $q_L$ and $Q_R^1$. With such a field content one could label $t_R$ and $q_L$ as ``elementary", with all other states labeled as ``composite".  The field content just described is summarized in Table~\ref{tab:top}.
\begin{table}[t]
\centering
\begin{tabular}{c|c|c|c|c||c|c||c|c}
	&\multicolumn{4}{c||}{scalar constituents}
	&\multicolumn{4}{c}{}\\ [0.3em]
\hline
\rule{0mm}{5mm}
Fermion &$Q_L^1$&$Q_L^2$&$S_L$&$S_R$&$Q^1_R$&$Q_R^2$
	&$q_L$&$t_R$\\ [0.3em]
\hline
\rule{0mm}{5mm}
$G=SO(5)_L \times U(1)_X$ 
	&\multicolumn{3}{c|}{$\bf 5_\frac{2}{3}$}&$\bf 1_\frac{2}{3}$
	&\multicolumn{2}{c||}{-}
	&-&-\\ [0.3em]
\hline
\rule{0mm}{5mm}
$H=SO(4) \times U(1)_X$
	&\multicolumn{2}{c|}{$(\bf 2, 2)_\frac{2}{3}$}&$\bf 1_\frac{2}{3}$&$1_\frac{2}{3}$
	&\multicolumn{2}{c||}{$\bf (2,2)_\frac{2}{3}$}
	&-&-\\ [0.3em]
\hline
\rule{0mm}{5mm}
$SU(2)_L\times U(1)_Y$
	&$\bf 2_\frac{1}{6}$&$\bf 2_\frac{7}{6}$&$\bf 1_\frac{2}{3}$&$\bf 1_\frac{2}{3}$
	&$\bf 2_\frac{1}{6}$&$\bf 2_\frac{7}{6}$
	&$\bf 2_\frac{1}{6}$&$\bf 1_\frac{2}{3}$\\ [0.3em]
\hline
\rule{0mm}{5mm}
	&\multicolumn{6}{c||}{``composite''}
	&\multicolumn{2}{c}{``elementary''}\\ [0.3em]
\end{tabular}
\caption{Quantum numbers of the top sector of the model, and its
relation to the PC picture.  The RH composite sector can be made fully
$G$ symmetric by adding another vector-like $H$ singlet (see main text).  The state
$(q_L,Q^1_R)$ can be decoupled by a large mixing mass
$\mu_{qQ}$.  The ``scalar constituents" are those that lead to light
scalar bound states.}
\label{tab:top}
\end{table}
The composite sector allows now for a global symmetry $SO(4)_L \times SO(4)_R$, where $(Q_L^1, Q_L^2)$ transform as $\bf (4,1)$, $(Q_R^1, Q_R^2)$ transform as $\bf (1,4)$ and all other fields transform as singlets. This symmetry is broken softly by mass terms such as Eq.~(\ref{mass1}) or mass terms of the form:
\be
\mathcal L_F^{mass,2}~=~-\mu_{QQ} \, \bar Q_L^1Q_R^1-\mu_{qQ} \, \bar q_LQ_R^1~.
\label{mass2}
\ee
For $\mu_{QQ}=\mu_{QQ}'$, the composite sector has an exact custodial $H \equiv SO(4) = [SO(4)_L \times SO(4)_R]_{\rm diagonal}$ global symmetry, that is only broken by mixing with the elementary fields $q_L$ and $t_R$. Taking $\mu_{QQ} \neq \mu_{QQ}'$ then corresponds to (soft) custodial breaking in the composite sector, which as we will see can be of phenomenological importance. Note that Eqs.~(\ref{singletmass}), (\ref{mass1}) and (\ref{mass2}) are the most general mass terms consistent with $SU(2)_L \times U(1)_Y$, given the field content of Table~\ref{tab:top}.

The approximate spectrum of the ``extended" model is then given by Eqs.~(\ref{massesf}) but with the top mass modified to
\be
m_t^2~=~\frac{s_v^2}{2}\frac{\xi^2\hat f^2\,\mu_{tS}^2\mu_{qQ}^2}{m_S^2m_Q^2}~,
\label{masstop}
\ee
[due to the additional masses in Eq.~(\ref{mass2})], together with an additional state with approximate mass
\be
m_Q^2~=~\mu_{QQ}^2+\mu_{qQ}^2~.
\label{massesf2}
\ee
The minimal model discussed at the beginning of this section can be recovered in the limit $\mu_{qQ}\to\infty$, and in PC language it features a ``mostly composite LH top''. Note that one can make the connection to other CH models even sharper by adding a vector-like $SU(2)_L$ singlet $(S'_L, S'_R)$ so that $\Psi_L = (Q^1_L, Q^2_L,S_L)$ and $\Psi_R = (Q^1_R, Q^2_R,S'_R)$ transform as a vectorlike $\bf 5$ of $SO(5)$, with $(S_R, S'_L)$ a vector-like $SO(5)$ singlet: in this scenario the ``composite sector" is explicitly $SO(5)$ [and not only $SO(4)$] invariant, the symmetry being reduced from $SU(5)$ down to $SO(5)$ due to the 4-fermion interactions discussed in the previous section. Writing a mass term for $(S'_L, S'_R)$ breaks this $SO(5)$ symmetry softly and allows to decouple these additional states. We therefore see that the minimal or extended models can be obtained from a softly broken $SO(5)$ invariant composite sector in an appropriate limit, which still leaves the $SO(5)_L$ symmetry discussed in Subsection~\ref{NJL} untouched. 

We will  examine the ``minimal" model with just the states $(Q^1_L, Q^2_L,
S_L)$, $S_R$, $Q^2_R$ and $t_R$, with the symmetry breaking Lagrangian
(\ref{mass1}), as well as the ``extended" model with the additional $Q_R^1,q_L$
and the symmetry breaking Lagrangian Eq.~(\ref{mass2}). While it is possible to achieve
realistic electroweak breaking within  the minimal model, we always
find a negative contribution to the $T$ parameter which makes
compatibility with EWPT challenging.  The extended model on the other
hand does not suffer from this problem. We will not analyze the ``fully $SO(5)$ invariant model" that includes also the states $(S'_L, S'_R)$ discussed above, since it does not introduce a qualitatively new feature compared to the extended model (which itself allows to introduce a controlled amount of custodial breaking that is not present in the minimal model).  Before analyzing these points in more detail, we will discuss the spin-1 sector of the NJL scenario.

\subsection{The Spin-1 Sector}
\label{sec:spin1}

In this section we turn to the composite vector resonances that are
predicted within our model.  The latter appear very naturally as
gauge-bosons of a hidden local symmetry \cite{Bando:1987br}.
Modeling the spin-1 resonances allows us to understand how the loop contributions to the (dynamically generated) Higgs potential are cut off by these heavy states.

One first notices that the NGB's defined by Eq.~(\ref{NGBExp}) can be
eliminated from the Yukawa couplings in
Eq.~(\ref{LSHLScanonicalNewBasis}) by a fermionic field redefinition of the form
\be
\bmat Q_L \\ S_L\emat ~=~U_{L}\bmat \Q_L\\\ \S_L\emat~,\qquad  
S_R~=~U_{R} \, \S_R~,
\label{HLSbasis}
\ee
with $Ue_5=U_Le_5U_{R}^\dagger$, and $Q_L = (Q^1_L, Q^2_L)^T$, $\Q_L = (\Q^1_L, \Q^2_L)^T$.
One obvious possibility is the choice
\be
U_{L} ~=~ U\,,\qquad 
U_{R} ~=~ 1\,.
\label{unitary}
\ee 
However, this choice is ambiguous up to a {\em local} $H_{\rm HLS}
\equiv SO(4) \times U(1)_X$ transformation that acts on $U_{L}$
and $U_{R}$ from the right.\footnote{In addition, the fields $U_{L,R}$ transform under the full global group $G$ acting from the left.}  This ambiguity defines a so-called
Hidden Local Symmetry (HLS) \cite{Bando:1987br} and
Eq.~(\ref{unitary}) corresponds to the unitary gauge.  Notice that
outside the unitary gauge, the matrix $U_{L}$ and the phase
$U_{R}$ also contain the NGB's of the HLS. 
It is furthermore convenient to parameterize $U_R$ and $U_L$ by $U_1$ and $U_5$ defined as \footnote{Notice that the $U(1)_X$ parts of $U_{L}$ and $U_{R}$ must coincide, as they transform identically under the Abelian part of $H_{\rm HLS}$, i.e.~formally we have  $[U_{L}^\dagger\partial_\mu U_{L}]^X=U_{R}^\dagger \partial_\mu U_{R}=\frac{2}{3} \, U_1^\dagger \partial_\mu U_1$.}
\be
U_R\equiv(U_1)^\frac{2}{3}\,,\qquad U_L\equiv U_5(U_1)^\frac{2}{3}~.
\ee
The field $U_1$ transforms under $U(1)_X$ with unit charge and the field $U_5$ is an element of $SO(5)$ only.

We will refer to the basis defined in Eq.~(\ref{HLSbasis}) as the HLS
basis and denote its fields with calligraphic letters.  Note that
these fields transform as \textit{singlets} of the global group $G$
but transform non-trivially under $H_{\rm HLS}$.\footnote{\label{footTop1} The construction of this section could be generalized to the ``full $SO(5)$ symmetric'' model discussed at the end of Section~\ref{top}, from which the minimal and extended models can be obtained after decoupling certain states. For instance, using the notation of Eq.~(\ref{HLSbasis}), one could rotate $(Q_R, S'_R)^T = U_L (\Q_R, \S'_R)^T$, while leaving the ``elementary'' $q_L$ and $t_R$ unchanged.
For
simplicity, we do not keep track of these fields in this section.} The
only fields that transform under $G$ are the NGB's in $U_{\bf 5}$ and
$U_{\bf 1}$.  After the above transformation, one obtains
\be
\mathcal L_F ~=~ i (\bar \Q_{L}, \bar \S_L) \gamma^\mu  \left(\partial_\mu+U_{5}^\dagger\partial_\mu U_{5}
+q_X U_{ 1}^\dagger\partial_\mu U_{1}
\right)\bmat \Q_L\\ \S_L\emat 
+i\bar\S_R \gamma^\mu \left(\partial_\mu+q_X U_{ 1}^\dagger\partial_\mu U_{1}\right)\S_R~,
\label{LFHLS}
\ee
where $q_X = \frac{2}{3}$, and
\be
\mathcal L_S ~=~ -\frac{1}{2 G_S}\H^2 - \H\, \bar\S\S~.
\label{LSHLS}
\ee
The Cartan connections appearing in $\mathcal L_F$ ensure that the
Lagrangian is fully invariant under the gauge symmetry $H_{\rm HLS}$.  We
will see now that they will become dynamical composite gauge fields,
in full analogy to the scalar composites above.

In addition to the scalar four-fermion channels, we add the
corresponding vector channels
\be
\mathcal L_{V}~=~-\frac{G_\rho}{2}
(J^{A\,\mu})^2 -\frac{G_X}{2}(J^{X\,\mu})^2~,
\label{currents}
\ee
with the conserved $SO(5)_L$ and $U(1)_X$
currents~\footnote{\label{footTop2} The normalization of the
generators is $\tr T^A T^B=\delta^{AB}$.}
\be
J^{A\,\mu}~=~(\bar Q_{L}, \bar S_L) \, T^A\gamma^\mu \bmat Q_L\\S_L\emat~,
\qquad
J^{X\,\mu}~=~q_X (\bar Q_L\gamma^\mu Q_L+\bar S_L\gamma^\mu S_L+\bar S_R\gamma^\mu S_R)~.
\label{currents2}
\ee
In full analogy to the scalar Lagrangian $\mathcal L_S$, we now
introduce spin-1 auxiliary fields
\be
\mathcal L_V~=~\frac{1}{2 G_\rho} (A^A_{\mu})^2+\frac{1}{2G_X} (A_\mu^X)^2+A_\mu^AJ^{A\mu}+A_\mu^XJ^{X\mu}~.
\label{auxiliarygauge}
\ee
In the HLS basis, Eq.~(\ref{HLSbasis}), it is then natural to define
new vector fields
\bea
\A_\mu^A&=&[U_{5}^\dagger(A_\mu+i\partial_\mu)U_{ 5}]^A~,
\nn\\
\A_\mu^X&=&A_\mu^X+iU_{1}^\dagger\partial_\mu U_{1}~.
\label{HLSgaugefields}
\eea
Indeed, one checks that the definitions Eq.~(\ref{HLSgaugefields}) are precisely the connections coupling to the fermions, and hence define their covariant
derivatives in the HLS basis [see Eqs.~(\ref{LFHLS}), (\ref{auxiliarygauge})]:
\be
D_\mu~=~\partial_\mu-i \, T^A\A_\mu^A  -i\, q_X \A_\mu^X~.
\label{Dmu}
\ee
Denoting the $SO(4)_L$-generators by $T^a$ and the $SO(5)_L/SO(4)_L$
generators by $T^{\hat a}$ (see App.~\ref{GroupTheory}), observe that
only $(\A^a_\mu,\A^X_\mu)$ transform as gauge fields of $H_{\rm HLS}$
while $\A^{\hat a}_\mu$ transform homogeneously.

At the UV matching scale, $\Lambda$, one ends up with the Lagrangian
\bea
\mathcal L_F+\mathcal L_S+\mathcal L_V&=&
i (\bar \Q_{L}, \bar \S_L) \Dsl \bmat \Q_L\\ \S_L\emat +i\bar\S_R \Dsl\S_R
-\frac{1}{2 G_S}\H^2- \H\, \bar\S\S
\nn\\
&&\mbox{} + \frac{1}{2G_\rho } \left( \A^A_\mu-i[ U_{5}^\dagger \partial_\mu U_{5}]^A \right)^2 
+\frac{1}{2G_X} \left( \A^X_\mu-i\, U_{1}^\dagger \partial_\mu U_{1}  \right)^2~.
\label{L1}
\eea

The gauging of the EW subgroup of $G$ proceeds by the
substitution $\partial_\mu U_{1,5}\to D^{SM}_\mu U_{1,5}$ (no
other fields transforms under $G$), with
\bea
D^{SM}_\mu U_{ 5} &=& \left[ \partial_\mu - i w^i_{L \mu} T^i_L - i b_\mu T^3_R  \right] U_{ 5}~,
\nonumber \\ [0.3em]
D^{SM}_\mu U_{ 1} &=& \left( \partial_\mu - i b_\mu \right) U_{1}~,
\label{DSM}
\eea
where $T^i_L$ (i = 1,2,3) are the $SU(2)_L$ generators and $T^3_R$ is the
third isospin generator of $SU(2)_R$ [see
Eq.~(\ref{OurBasis})].  One also introduces the kinetic terms
\bea
\mathcal L_{G} &=& -\frac{1}{4g_0^2}(w_{\mu\nu}^a)^2 - \frac{1}{4g'^2_0}(b_{\mu\nu})^2~.
\label{L2}
\eea
RG running induces kinetic terms for $\A_\mu$ and $\H$,
\bea
\mathcal L_K &=& -\frac{1}{4g_\rho^2}(\F^V_{\mu\nu})^2-\frac{1}{4g_X^2}(\F^X_{\mu\nu})^2-\frac{1}{2}(\nabla_\mu\H)^2~,
\label{L3}
\eea
with
\bea
\nabla_\mu \H &=& (\partial_\mu-i \A_\mu^{\hat a} \, T^{\hat a}) e_5 \H~,
\label{delmu}
\eea
and where we normalized canonically the field $\H$ in full analogy to
Eqs.~(\ref{LSHLScanonical}) or (\ref{LSHLScanonicalNewBasis}), but
kept a convenient non-canonical normalization for the spin-1 fields.

Putting all the ingredients together, the Lagrangian of the model reads
\bea
\mathcal L &=& i (\bar \Q_{L}, \bar \S_L) \Dsl \bmat \Q_L\\ \S_L\emat +i\bar\S_R \Dsl\S_R
+ \frac{1}{2}(\nabla_\mu\H)^2 - \frac{1}{4} \lambda \left( \H^2 - \hat{f}^2 \right)^2 - \xi \, \H\, \bar\S\S
\nonumber \\ [0.3em]
& & 
\mbox{} + \frac{1}{4} f^2_\rho \left( \A^A_\mu-i[ U_{5}^\dagger  D^{SM}_\mu U_{ 5}]^A \right)^2 
+\frac{1}{4} f^2_X \left( \A^X_\mu-i\, U_{1}^\dagger  D^{SM}_\mu U_{1}  \right)^2
\label{LFinal}
\\ [0.3em]
& & 
\mbox{} - \frac{1}{4g_\rho^2}(\F^V_{\mu\nu})^2-\frac{1}{4g_X^2}(\F^X_{\mu\nu})^2
-\frac{1}{4g_0^2}(w_{\mu\nu}^a)^2 - \frac{1}{4g'^2_0}(b_{\mu\nu})^2~,
\nonumber
\eea
together with Eqs.~(\ref{Dmu}), (\ref{DSM}) and
(\ref{delmu}).\footnote{It only remains to add terms associated with
the fermionic extension as well as the soft-breaking terms in
Eqs.~(\ref{mass1}) or (\ref{mass2}), as discussed in Section~\ref{top} and
Footnotes~\ref{footTop1} and \ref{footTop2}.} For later convenience,
we introduced the decay constants $f_\rho^2=2G_\rho^{-1}$ and $f_X^2=2G_X^{-1}$ of the
$SO(5)_L$ and $U(1)_X$ resonances, respectively.  The physical NGB decay constant $f$ is then
obtained after integrating out the coset resonances and is given by
\bea
f^{-2} = \hat f^{-2}+f_\rho^{-2}~.
\eea
The physical spin-1 masses (before EWSB) are given by
\be
m_\rho^2 ~=~ \frac{g_\rho^2f_\rho^2}{2}\,,\qquad 
m_a^2~=~
r_v^{-1}\, m_\rho^2~,\qquad  
m_X^2 ~=~ \frac{g_X^2f_X^2}{2}\,,
\qquad 
\textrm{where}
\qquad
r_v\equiv \frac{f^2}{\hat f^2}<1~,
\label{vectormasses}
\ee
where the index $\rho$ denotes the $SO(4)_L$ resonances and the index
$a$ the $SO(5)_L/SO(4)_L$ ones. The remaining spin-1 states are massless in this approximation and can be identified with the SM gauge bosons. The corresponding gauge couplings are given in terms of the fundamental parameters by $1/g^2 = 1/g_0^2 + 1/g_\rho^2$ and $1/g'^2 = 1/g_0'^2 + 1/g_\rho^2 + 1/g_X^2$.

In a large part of the literature on NJL models, the couplings
$g_\rho$ and $\xi$ are studied in the so-called fermion loop
approximation, in which only the planar fermion loops -- leading in
the number of fermion colors $N_c$ -- are kept \cite{Wakamatsu:1988ht}.  In this case,
the beta functions are one-loop only, and in particular are positive,
hence rapidly decreasing the couplings $\xi,g_\rho$ in the IR. This
allows one to conclude that the compositeness boundary condition
$\xi,g_\rho=4\pi$ at a UV scale $\Lambda$ can indeed be consistent
with relatively weakly coupled states $g_\rho,\xi\ll 4\pi$ with masses
below the compositeness scale $\Lambda$.  The real-world value of the
number of colors being $N_c=3$, the validity of this approximation is
far from obvious.

As it turns out, there is a fundamental difference in the RG running
of the nonabelian coupling $g_\rho$ and the Yukawa coupling $\xi$ once
the diagrams subleading in $N_c$ are included (such as loops of the
actual composite states $\H$ and $\A_V$).  While the full 1-loop
contribution to $\beta_\xi$ remains positive, the one for
$\beta_{g_\rho}$ switches sign due to the negative contribution of the
gauge self interactions, $g_\rho$, at least for $N_c=3$ and large
enough global groups such as $SO(5)$.  Since beyond the large $N_c$
approximation we would in principle need to take into account higher
loop contributions (at least near $\Lambda$ where $\xi,g_\rho\sim
4\pi$), one cannot reach any safe conclusion as to whether any
composite states $\A^V_\mu$ below $\Lambda$ are present or not.

To shed more light on this issue, it is useful to imagine there was
such a relatively weakly coupled state $g_\rho\ll 4\pi$ with mass
$m_V\ll \Lambda$.  Then the $\beta$ function can safely be
approximated by its one-loop value, and evolving the coupling $g_\rho$
towards the UV will further decrease it.  One would tend to conclude
that a compositeness boundary condition $g_\rho=4\pi$ can never be
reached.  However, the Yukawa couplings will start increasing and will
eventually impact the running of $g_\rho$ via higher-loop effects.
Similarly, the NGB that make up the longitudinal components of the
spin-one resonances are expected to become strongly coupled in the UV,
triggering a breakdown of perturbative unitarity.  Although this is
expected to happen only near the compositeness scale (that we might
define as $\xi(\Lambda)=4\pi$ in this case) it cannot be said with
certainty if these effects are able to sufficiently increase $g_\rho$ to allow for a compositeness boundary condition at such scales.

In conclusion, the IR value for $g_\rho$ -- unlike the coupling $\xi$
-- cannot be predicted.  We will leave it as a completely free
parameter of the model, keeping in mind the possibility that the
composite states $\A^V$ could be strongly coupled with a mass near the
cutoff.\footnote{It is worth noticing that the running of the Abelian
coupling $g_X$ does not suffer from this issue, and can be computed
reliably.} We will see that unless $g_\rho$ is very close to $4\pi$,
our predictions depend very little on its precise value.  In practice
we can simply ignore the RG running of all couplings in the UV and
trade the value of the UV scale $\Lambda$ for the the IR value of
$\xi$.  The scalar self coupling $\lambda$ remains predicted due to
the IR quasi-fixed point. 

Before closing this section, we comment on two further aspects directly connected to the spin-1 sector.

\medskip
\noindent
\textbf{Connection to 2-site models} \\
Interestingly, our Lagrangian Eq.~(\ref{LFinal}) is equivalent to a
two-site model.  Recall that a two-site model is defined as follows.
The first site has global group $G_0=SO(5)$ of which only the SM
subgroup is gauged.\footnote{For simplicity, we omit in this
discussion the $U(1)_X$ factors.} On the second site one introduces a
CCWZ \cite{Coleman:1969sm,Callan:1969sn} type breaking of the group
$G_1=SO(5)$ to $H_1=SO(4)$, parameterized by a field $\Phi$.  The two
sites interact via \textit{link fields}, $\Omega$, which transform in
the bi-fundamental of $G_0\times G_1$.  Focusing on the spin-1 part,
the Lagrangian is thus given by \cite{Panico:2011pw,DeCurtis:2011yx}
\bea
\mathcal L &=& -\frac{1}{4g_0^2}\tr (F^0_{\mu\nu})^2-\frac{1}{4g_1^2}\tr(F^1_{\mu\nu})^2+ \frac{f_1^2}{2}\tr D_\mu\Omega^\dagger D_\mu\Omega+ \frac{1}{2}|D_\mu \Phi|^2~.
\label{2site}
\eea
The covariant derivatives are
\bea
D_\mu \Omega&=&\partial_\mu\Omega-iA_\mu^0\Omega+i\Omega A_\mu^1~, \\
D_\mu \Phi &=&\partial_\mu \Phi-iA_\mu^1\Phi~,
\eea 
where $A_0$ are the ``elementary" gauge fields and $A_1$ the
composite ones.  In the last term of Eq.~(\ref{2site}), $\Phi=\hat f
\tilde U e_5$ accomplishes the breaking $G_1\to H_1$.  The matrix $\tilde
U$ contains the Goldstone bosons of this breaking on the
second site, while the link field $\Omega$ contains the NGB of the
breaking to $G=(G_0\times G_1)_{\rm diagonal}$.  Going to the gauge
$\tilde{U}=1$ we precisely recover our Lagrangian with the identifications
\be
\Omega~=~U_{\bf 5}\,,\qquad 
f_1^2~=~ f_\rho^2 \,,\qquad 
g_1~=~g_\rho\,.
\ee

Finally, we point out that the Lagrangian of Eq.~(\ref{2site}) (and
hence the spin-1 sector of our model) is actually equivalent to the
most general, left-right symmetric Lagrangian of one complete set of
$G$ resonances, if and only if the spin-1 sum rules in Eqs.~(3.13) and (3.14) of \cite{Marzocca:2012zn} hold.

\medskip
\noindent
\textbf{Tree-level $S$-parameter from the vector resonances} \\
One of the most sensitive constraints on any composite Higgs model
comes from the electroweak $S$ parameter, and its largest contribution
is provided by the spin-1 resonances.  Since our spin-1 sector is
quite general and basically the same as in any composite Higgs model,
it is worthwhile to consider at this stage the constraints implied by
$S$.  It can be computed as (see App.~\ref{app:potential})
\be
S=-16 \pi \Pi_{3B}'(0)=4\pi s_v^2 f^2\left(\frac{1}{m_\rho^2}+\frac{1}{m_a^2}\right)=8\pi s_v^2 \frac{1-r_v^2}{g_\rho^2}
\ee
Using the most recent oblique fit (with $U=0$) \cite{Baak:2014ora}
\be
S=0.06\pm 0.09\,,\qquad T=0.1\pm 0.07
\ee
with a correlation of $\rho=0.91$, one can place bounds on the
parameters.  At $T=0$ the 95\% C.L.~interval for $S$ is
$[-0.13,0.017]$, while the SM point $S=T=0$ roughly sits on the 90\%
C.L. contour.  For $r_v=0.5$ we find that $g_\rho=3\ (6)$ requires
$f\gtrsim 2.7\ (1.3)$ TeV.  These bounds can be alleviated (made worse) if quantum corrections
yield negative (positive) contributions to $S$ or positive (negative) contributions to $T$.

\subsection{Parameter Space of the Model}
\label{sec:pars}

The parameter space of the model is spanned by the couplings $\xi$,
$g_\rho$, the symmetry breaking VEV $\hat f$, the NGB decay constant
$f$ and the fermionic mass parameters $\mu_{QQ}$, $\mu'_{QQ}$,
$\mu_{tS}$ and $\mu_{qQ}$.\footnote{ We will analyze first the case
where the tadpole term $\tau$ vanishes, and only include this
parameter at a second stage in our analysis. Also, as we show in App.~\ref{app:vector}, when $g^2_X \gg g_0'^2$ the $U(1)_X$ spin-1 resonance induces only minor effects.} These are eight
parameters that can be expressed in terms of the masses $m_{S}$,
$m_Q$, $m'_{Q}$, $m_\rho$, $m_a$, and $m_t$ and two mixing angles:
\be
s_R\equiv \frac{\mu_{tS}}{m_S}\,,\qquad s_L\equiv \frac{\mu_{qQ}}{m_Q}\,,\qquad s_{L,R}=\sin \alpha_{L,R}~.
\label{sLsR}
\ee
The mass of the radial mode can be expressed in terms of these
parameters as
\be
m_\H = \sqrt{2 a_*}\, c_R\, m_S~.
\ee
Since we have two conditions from electroweak breaking, we can
eliminate the angles $s_L$, $s_R$, and the radial mass will be a
prediction in terms of the other masses alone.  Moreover, the
electroweak splitting for the doublets will also be predicted.

Note that the mass condition for the top quark in Eq.~(\ref{masstop})
in this parametrization reads
\be
m_t = \frac{m_S}{\sqrt{2}} s_v s_L s_R c_R~,
\ee
which in particular implies the inequality
\be
m_S\geq 2\,y_tf~,
\label{ineq}
\ee
where we used that the gauge boson masses determine $s_v={v_{\rm
SM}}/{f}$ with $v_{\rm SM}=246$ GeV.

We will often use the notation
\be
r_v~\equiv~\frac{m_\rho^2}{m_a^2}\,,\qquad r_f~\equiv~\frac{m_Q^2}{m_S^2}\,,\qquad r_f'~\equiv~\frac{m_Q'^2}{m_S^2}~,
\ee
to denote ratios of the various masses.

%
%

\section{Remarks about the pNGB Effective Potential}
\label{GBpot}

One of the interesting features of the previous construction is that the Higgs boson appears quite explicitly as a bound state of fermions, and that its potential is generated dynamically. Since the Higgs is a pNGB of the breaking $SO(5) \to SO(4)$, the generated potential must be proportional to the couplings that break $SO(5)$ explicitly. In the spin-1 sector, the breaking arises from the gauging of the SM subgroup (hence the effects are proportional to $g$ and $g'$). In the fermionic sector, the parameters that break the $SO(5)$ symmetry are $\mu_{tS}$, $\mu_{QQ}$ and $\mu'_{QQ}$.\footnote{In the extended model, there can also exist contributions proportional to $\mu_{qQ}$. These are finite.} Note that the explicit breaking in the fermionic sector is soft. As we will see, at 1-loop order, the dominant effects are proportional to a certain combination of the previous parameters, such that when $\mu_{tS} = \mu_{QQ} = \mu'_{QQ}$ and the $SO(5)$ symmetry is restored, the corresponding contributions vanish. One attractive feature of many recent pNGB Higgs constructions is that the explicit breaking of the underlying symmetry whose spontaneous breaking leads to the pNGB's is ``super-soft", i.e.~it causes the Higgs potential to depend little or not at all on the details of the UV completion, resulting in a completely predictive model of EWSB.\footnote{In N-site models, the potential is finite up to a fixed order in the loop expansion. Often, sensitivity to the UV physics at the scale $\Lambda$ can be introduced at sufficiently high order. In extra-dimensional construction the Higgs potential is finite due to locality in the extra-dimension.} We will see that this holds true in our case, due to an interesting twist that is due to the RG running above the symmetry breaking scale.

Due to the shift-symmetry, the effective potential depends on the angular variable $s_h = \sin(h/f)$. We give general formulas and analytic approximations for our setup in App.~\ref{app:potential}. In this section we simply highlight the main ingredients. If one formally expands the potential in powers of $s_h$, one has the parameterization
\bea
V &=& -\frac{\alpha}{2}s_h^2+\frac{\beta}{4}s_h^4 + {\cal O}(s_h^6)~,
\label{alphabeta}
\eea
where, as mentioned above, the explicit breaking of the global $SO(5)$ symmetry generates nonzero values for $\alpha$ and $\beta$. As discussed in App.~\ref{app:IRRegulation}, the naive expansion of the potential in powers of $s_h$ introduces a logarithmic IR divergence in $\beta$. A more careful analysis shows that this divergence is roughly cut by the $W$ or top masses, but the details of how this happens are not very important for the following discussion.

\begin{figure}
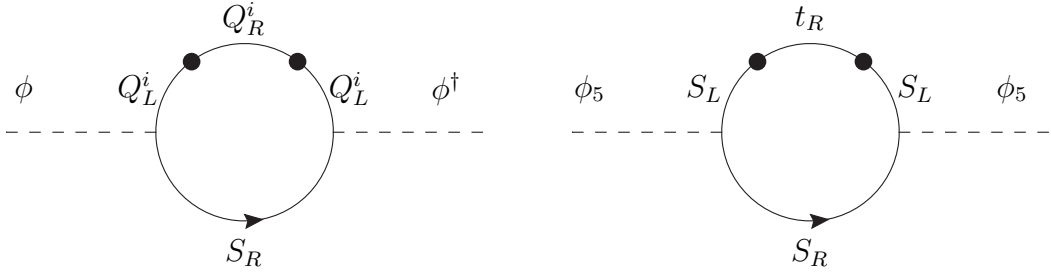

\centering
\psfrag{a}[c]{$\phi$}
\psfrag{b}[c]{$Q^i_L$}
\psfrag{c}[c]{$Q^i_R$}
\psfrag{d}[c]{$Q^i_L$}
\psfrag{e}[c]{$\phi^\dagger$}
\psfrag{f}[c]{$S_R$}
\psfragfig[width=0.4\linewidth]{Figures/diag1.eps}
\hspace{1cm}
\psfrag{a}[c]{$\phi_5$}
\psfrag{b}[c]{$S_L$}
\psfrag{c}[c]{$t_R$}
\psfrag{d}[c]{$S_L$}
\psfrag{e}[c]{$\phi_5$}
\psfrag{f}[c]{$S_R$}
\psfragfig[width=0.4\linewidth]{Figures/diag1.eps}
\caption{Logarithmically divergent diagrams contributing to the running of the operators $2|\phi|^2=\H^2s_h^2$ (left) and $\phi_5^2=\H^2c_h^2$ (right).}
\label{feynman1}
\end{figure}

We start with the contributions to  $\alpha$ and $\beta$ from fermion loops.
As it turns out, the top sector introduced in Section~\ref{top} is not enough to render the top-loop contribution to the NGB potential completely finite. However, the UV sensitivity is softened and only a logarithmic divergence of the Higgs mass remains. This fact can easily be understood as follows. 
Let us parameterize a $SO(5)_L$ violating mass splitting in the Higgs potential as
\bea
\mathcal L_{\rm mass} &=& -\frac{1}{2}m_1^2|\phi_5|^2-m_4^2|\phi^2|=-\frac{1}{2}m_1^2\bigl(|\phi_5|^2+2|\phi|^2\bigr)-\delta m^2\, |\phi|^2
\eea
The renormalization of the universal $SO(5)_L$ symmetric mass operator $\phi_5^2+2|\phi|^2=\H^2$ is quadratically divergent, but contains no NGB's. To renormalize the $SO(5)_L$ violating (NGB-dependent) mass operator $2|\phi|^2=\H^2 s_h^2$, one needs two fermionic, $SO(5)_L$ violating mass insertions as shown in Fig.~\ref{feynman1}, thus reducing  the divergence to a logarithmic one. By an analogous argument one can easily see that the one-loop renormalization of operators quartic in  $\phi$ and $\phi_5$ other than the $SO(5)_L$ symmetric one are completely finite. The logarithmic divergence for $\delta m^2$ is proportional to the combination of mass insertions 
\be 
\muext^2\equiv 2\mu_{tS}^2-\mu_{QQ}^2-\mu_{QQ}'^2~,
\label{mueff}
\ee
where the various masses were defined in Eqs.~(\ref{mass1}) and (\ref{mass2}).
A vanishing of this quantity would indicate a softer (actually finite) UV behavior, similar to the general sum-rules found in Ref.~\cite{Marzocca:2012zn}. However,   we will {\em not} assume $\muext=0$ in this paper. As a consequence, we are forced to introduce a counterterm for the mass splitting $\delta m^2$.
Once such a counterterm is introduced, it also receives multiplicative (logarithmically divergent) renormalization from scalar loops.

\begin{figure}
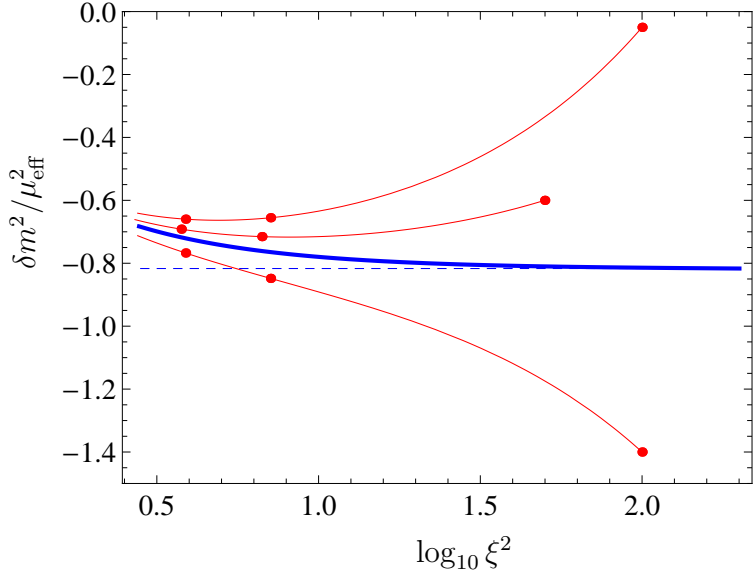

\centering
\psfrag{xxx}{$\log_{10} \xi^2$}
\psfrag{yyy}{$\delta m^2/\mu_{\rm eff}^2$}
\psfragfig[width=0.6\linewidth]{Figures/RGplot2}
\caption{RG flow of the ratio $\delta m^2/\mu_{\rm eff}^2$ and $\xi$. The dashed line marks the exact IR fixed point $-r_*$ that is reached in the absence of gauge interactions, the solid blue line is the asymptotic trajectory including QCD effects. The dots on the trajectories represent e-folds of RG running.}
\label{RGplot2}
\end{figure}

A major point that we would like to stress is that, remarkably, the introduction of a counterterm for the mass squared of the Goldstone boson does not reduce the predictivity of our model. The reason is similar to the prediction of the quartic self-coupling in terms of the Yukawa coupling that we encountered in Section~\ref{NJL}.
In fact, in the absence of gauge interactions,  the RG-equations for $\mu_{\rm eff}$ and the NGB mass $\delta m^2$ are
\be
\beta_{\muext^2} ~=~ \frac{\xi^2}{16\pi^2} \, \muext^2~,\qquad 
\beta_{\delta m^2} ~=~ \frac{3\xi^2}{4\pi^2} \, \muext^2+\frac{\lambda+3\xi^2}{4\pi^2} \, \delta m^2~.
\label{betamass}
\ee
The coupling $\mu_{\rm eff}$ only runs due to the anomalous dimension of the left handed fields, Eq.~(\ref{gammas}), while $\delta m^2$ contains the above mentioned logarithmically divergent fermion and scalar contributions. 
The crucial observation is that the system in Eq.~(\ref{betamass}) has a fixed point at
\be
\delta{m^2}~=~-r_*\mu_{\rm eff}^2~,
\label{FP2}
\ee
with $r_*\equiv \frac{156}{191}\approx 0.81$. This fixed point is IR stable and hence any unknown UV value for $\delta m^2$ is eliminated along the RG flow.
%
%
As was the case with the ratio $\lambda/\xi^2$, QCD corrections induce a mild dependence of $r_*$ on $\xi$.~\footnote{In addition to the modification of the running of the Yukawa coupling, there is a correction to the $\beta$ function for $\mu_{\rm eff}$ (see App.~\ref{app:RGeqs}).} This effect is illustrated in Fig.~\ref{RGplot2}, and  
results in slightly smaller values of $r_*$, e.g.~$r_*|_{\xi=2}\approx 0.72$.
One also observes that 
the asymptotic line is reached more slowly than in the case of the quartic coupling. 

The RG flow discussed so far applies to the regime above the masses of the fermionic and scalar resonances. The full effective potential is computed in 
App.~\ref{app:potential} which is used for our numerical analysis in Section~\ref{sec:ewsb}.
We can cross-check our claims above with this explicit calculation by combining the scalar and fermion loop contributions, and substituting the fixed-point conditions Eqs.~(\ref{FP1}) and (\ref{FP2}), obtaining
\be
\alpha_{0}+\alpha_{1/2}=r_*\mu_{\rm eff}^2\hat f^2\left(1+
\frac{11}{16\pi^2}\, \xi^2\,
\log\frac{M}{m_\H}\right) +{\ \rm finite}~,
\label{alphafermionscalar}
\ee
where $M$ is the renormalization scale. For the purpose of this discussion, we have chosen the IR cutoff as $m_\H$, the mass of the radial scalar mode [see Eq.~(\ref{massH})], assuming that the remaining resonances are somewhat lighter and therefore contribute to the running with full strength above $m_\H$.\footnote{A more precise treatment will identify the relevant thresholds and integrate out the respective degrees of freedom accordingly in order to identify the pNGB mass parameter at the weak scale. In this work we will be satisfied with the leading approximation that does not take into account such subtleties.} The first term thus corresponds to $-\delta m^2 \hat{f}^2$ at the IR quasi-fixed point discussed above, while the second accounts for the running between $M$ and $m_\H$. One should note that in reaching the quasi-fixed point value, the explicit loop suppression factor is lost, and the size of this contribution to the pNGB Higgs mass parameter is controlled by $\mu_{\rm eff}^2$.
We should find that the explicit RG scale-dependence in Eq.~(\ref{alphafermionscalar}) precisely accounts for the running of $\mu_{\rm eff}^2$ 
and the field rescaling of $\H$.
Indeed, from Eq.~(\ref{betamass}) and Eq.~(\ref{gammas}) one finds
\be
 \frac{\beta_{\mu_{\rm eff}^2}}{\mu_{\rm eff}^2}-2\gamma^{\ }_{\Phi}=-\frac{11}{16\pi^2}\,\xi^2.
\ee
$M$ can be chosen at will as long as the running parameters are evaluated at that scale. It is thus natural to chose  $M \sim m_\H$, in which case the parenthesis in  Eq.~(\ref{alphafermionscalar}) is close to one. 
The finite pieces of $\alpha$ and the 
corresponding contribution to $\beta$ are given in App.~\ref{app:potential}.

We remark that the modification needed for the minimal model (in which the hypercharge $1/6$ resonance is decoupled) is simply to replace $\muext$ by 
\be
 \mumin^2\equiv 2\mu_{tS}^2-\mu_{QQ}'^2~,
\label{mumin}
\ee
whereas Eq.~(\ref{betamass}) and hence the value of $r_*$ remain  unchanged, as the decoupled state $(q_L,Q_R^1)$ does not possess any Yukawa interactions.

 Finally, we comment on the contribution from the spin-1 sector, derived in Sec.~\ref{sec:spin1}. Details of the computation can be found in App.~\ref{app:potential}. Here we stress the fact that the resulting contributions to both $\alpha$ and $\beta$ are UV finite (cutoff at $\sim m_\rho$), due to the UV behaviour of the form factors Eq.~(\ref{spin1formfactor}), which are derived from the Lagrangian (\ref{LFinal}) or the equivalent 2-site Lagrangian (\ref{2site}).
As discussed in Sec.~\ref{sec:spin1}, we are assuming that a kinetic term for the spin-1 resonances is generated in the IR, which can be verified (and computed explicitely) in the large $N_c$ limit, but is not completely straightforward beyond that approximation.

\section{Electroweak Symmetry Breaking and the Higgs Mass}
\label{sec:ewsb}

Our starting point is the parametrization of the Higgs potential in the approximation of Eq.~(\ref{alphabeta}). EWSB and the correct Higgs mass imply the following simple conditions:
\be
\alpha~=~\frac{m_h^2f^2}{2c_v^2}~\approx~\frac{(88{\ \rm GeV})^2\, f^2}{c_v^2}\,,\qquad 
\beta~=~\frac{m_h^2 f^2}{2c_v^2s_v^2}~\approx~\frac{0.13\, f^4}{c_v^2},
\label{ewsb}
\ee
%
Notice that the Higgs VEV $v = \langle h \rangle$ is not equal to the SM model value $v_{\rm SM}=246$ GeV but is rather fixed by the relation $s_v f= v_{\rm SM}$.

A crucial observation that will facilitate the discussion below is that one has to expect a certain degree of cancellation between the different contributions to Eq.~(\ref{mueff}).
Observe that the leading contribution to $\alpha$ arises from Eq.~(\ref{alphafermionscalar}) and is given by
\be
\frac{\alpha}{f^2}~=~r_*\muext^2r_v^{-1} + \dots \,,
\label{alphatree}
\ee
The parameter $\mu'_{QQ}$ equals the mass $m'_Q$ of the exotic doublet $Q^2$ and it is already constrained by direct LHC searches of strongly pair-produced $Q=5/3$ fermions to be larger than about $800$~GeV at 95\%~C.L.~\cite{Chatrchyan:2013wfa}.\footnote{The lower limits for $Q = 2/3$ top-partners are around $700-800~{\rm GeV}$, depending on the decay mode~\cite{Chatrchyan:2013uxa,ATLASTopPartner8TeVCONF}, while $Q=-1/3$ fermionic resonances should be heavier than about $500-800~{\rm GeV}$~\cite{CMSBSearch}. See also~\cite{Gripaios:2014pqa}.} However, if $\muext$ was of that size, then its contribution would need to be canceled by other (truly loop-suppressed) contributions, which requires large couplings $g_{\rho}, \xi$, possibly outside the perturbative regime. One thus expects a certain degree of cancellation to happen between the different terms in $\muext^2$ such that~\footnote{Notice that as $\mu_{\rm eff}$ renormalizes only multiplicatively this is a technically natural tuning. We come back to the issue of naturalness in Section~\ref{naturalness}.}
\be
|\epsilon| \ll 1\,,\qquad \textrm{where} \qquad \epsilon\equiv\frac{\muext^2}{2\mu_{tS}^2}~.
\label{mueff2}
\ee
To reasonably good approximation one can thus set $\muext\approx 0$ everywhere but in the leading contribution to 
$\alpha_{1/2}$ given by Eq.(\ref{alphatree}) (for instance, we can approximate $\beta_{1/2}\approx \beta_{1/2}|_{\muext=0}$).
In the minimal model, the same argument leads one to conclude that
\be
|\tilde\epsilon| \ll 1\,,\qquad \textrm{where} \qquad \tilde\epsilon\equiv\frac{\mumin^2}{2\mu_{tS}^2}~.
\label{mumin2}
\ee
We will make use of these relations in the discussions below, but will always keep the exact expressions in our numerical analyses.

We will start our analysis with the minimal scenario described in Section~\ref{top}, and first consider the case with vanishing tadpole. 
Using then Eq.~(\ref{mumin2}) in the expression for the (dominant) fermionic contribution to $\beta$, one obtains 
\bea
\frac{\beta_{1/2}}{f^4} &=& \frac{3}{4\pi^2}\frac{m_t^4}{v_{\rm SM}^4}\left(
\log\frac{m_Q'^2}{m_t^2}-\frac{r_f'^2(3-r'_f)}{(1-r'_f)^3}\log r'_f-\frac{1+r_f'^2}{(1-r'_f)^2}
\right)\nn\\
&\leq&  \frac{3}{4\pi^2}\frac{m_t^4}{v_{\rm SM}^4}\left(
\log\frac{m_Q'^2}{m_t^2}-1
\right)
~.
\label{smallbeta}
\eea
where the upper bound in the second line is attained for $r_f'\to 0$.
This is too small  and, comparing to Eq.~(\ref{ewsb}), implies that the correct Higgs mass cannot be achieved unless $c_v \approx 1$ (e.g.~$f > 1~\rm{TeV}$) and $m_{Q'} \gtrsim 1.7~\rm{GeV}$ (using the exact expressions, the situation is actually worse, and typically we find too light a Higgs). 
A very similar conclusion is reached for the extended model with the $SO(4)$ symmetric choice $\mu_{QQ}=\mu_{QQ}'$. In this case Eq.~(\ref{mueff2}) reduces $\beta_{1/2}$ to the same expression, Eq.~(\ref{smallbeta}), with the replacement $r_f'\to r_f$, $m_Q'\to m_Q$ and hence
\be
\frac{\beta_{1/2}}{f^4}\leq  \frac{3}{4\pi^2}\frac{m_t^4}{v_{\rm SM}^4}\left(
\log\frac{m_Q^2}{m_t^2}-1\right)
~.
\label{smallbetaextended}
\ee

A possible  solution to this problem is to add the explicit $SO(5)$ breaking tadpole term, Eq.~(\ref{tadpole}), which easily accounts for the missing contribution to $\beta$. One obtains
\be
\alpha_\tau~=~-\tau\hat f~,\qquad
\beta_\tau~=~\frac{1}{2}\tau\hat f~.
\ee
As $\beta_{1/2}+\beta_1$ is positive but too small, $\tau$ needs to be positive but is bounded from Eq.~(\ref{ewsb}) by 
\be
\tau \lesssim 0.26\, \frac{f^4}{\hat f\, c_v^{2}} \,,
\ee
In particular, the correction to the mass of the radial mode due to the tadpole, cf.~Eq.~(\ref{massH2}), is  negligible. 
Moreover the relation (\ref{mueff2}) remains true, as $\alpha_\tau$ can never become large enough so as to significantly reduce the leading contribution to $\alpha$.

\begin{figure}
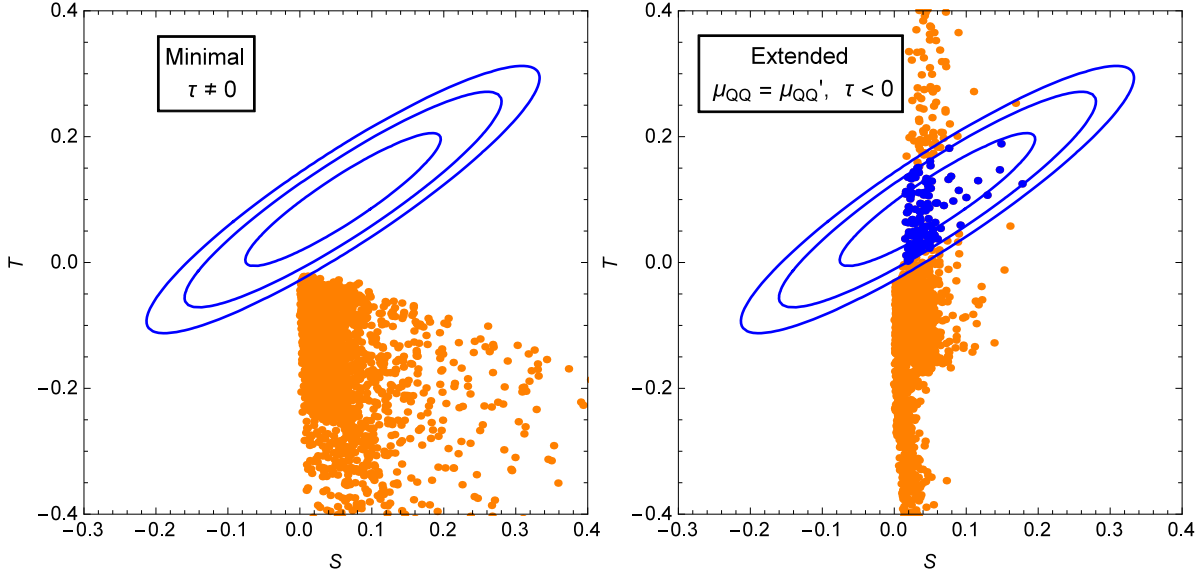

\psfrag{S}{$S$}
\psfrag{T}{$T$}
\psfragfig[width=0.48\linewidth]{Figures/STminimal}
\psfragfig[width=0.48\linewidth]{Figures/STextended0}
\caption{Electroweak precision tests for the minimal model with $\tau\neq0$ (left) and the extended model with $\mu_{QQ}=\mu_{QQ}'$ and $\tau<0$ (right). We scan over the ranges $f \in [500,2000]~\rm GeV$, $r_v \in [0.05, 0.95]$, $g_\rho \in [0,3\pi]$, $s_R \in [0,1]$ (and, for the right plot, $s_L \in [0,1]$). We fix $\tau$, $m_S$ and $m_Q$ from EWSB (Higgs vev and Higgs mass) plus the top mass, but requiring $m_S, m_Q > 500~\rm{GeV}$. In the left panel we also require $|\tau| \in [0,(1000~\rm{GeV})^3]$ while in the right panel we impose $\tau \in [-(3000~\rm{GeV})^3, 0]$. All points reproduce the correct Higgs, top and $Z$ masses. The contours correspond to 68\%, 95\% and 99\% C.L.~respectively~\cite{Baak:2014ora}.} 
\label{STminimal}
\end{figure}

However, as it turns out, the fermionic contribution to the $T$ parameter is negative in the minimal model, for reasons closely connected to those first discussed in~\cite{Carena:2006bn}. 
As we have already pointed out at the end of Section~\ref{sec:spin1}, this means that EWPT are very difficult to satisfy. In fact, performing a scan of the minimal model over 3000 points with $0.5$ TeV $<f<2$ TeV, we find not a single point within 95\% C.L., and only 20 points pass EWPT at 99\% C.L. (see the left panel of Fig.~\ref{STminimal}).\footnote{Here and in the following we also include an additional contribution to the $T$ parameter 
$\Delta T=-\frac{3}{8 \pi}\frac{m_Z^2}{m_W^2}\log\frac{\Lambda}{m_h}$
due to modified Higgs couplings \cite{Barbieri:2007bh}. For definiteness we use $\Lambda = m_\rho$ in our scans. }

A next to simplest model is the extended model with $\mu_{QQ}'=\mu_{QQ}$, and $\tau>0$ to raise the Higgs mass. This model has the interesting feature that the tuning Eq.~(\ref{mueff2}) is protected, as the  point $\mu_{QQ}=\mu_{QQ}'=\mu_{tS}$ is invariant under the global symmetry. However, as with the minimal scenario, for $\tau >0$ all points that lead to successful EWSB have a negative $T$ and do not pass EWPT. 
There exists however a possibility in the latter scenario to accommodate both the correct Higgs mass and EWPT with a large negative value for $\tau$. In fact, if $-\tau$ is so large as to cancel a large negative $\muext^2$, we can escape the condition (\ref{mueff2}) and $\beta_{1/2}$ is not bounded by (\ref{smallbetaextended}) . This will however require a very large $|\tau|$, and both  $\alpha$ and $\beta$ show substantial cancellations between tadpole and other contributions. Notice that a large negative $\tau$ is bounded by the mass for the radial mode, Eq.~(\ref{massH2}). We show the $S$ and $T$ parameters of this model in the right panel of Fig.~\ref{STminimal}. We find that the interplay of EWSB and EWPT require in this case a peculiar hierarchy of fermion masses, $m_S< m'_Q<m_Q$.

\begin{figure}
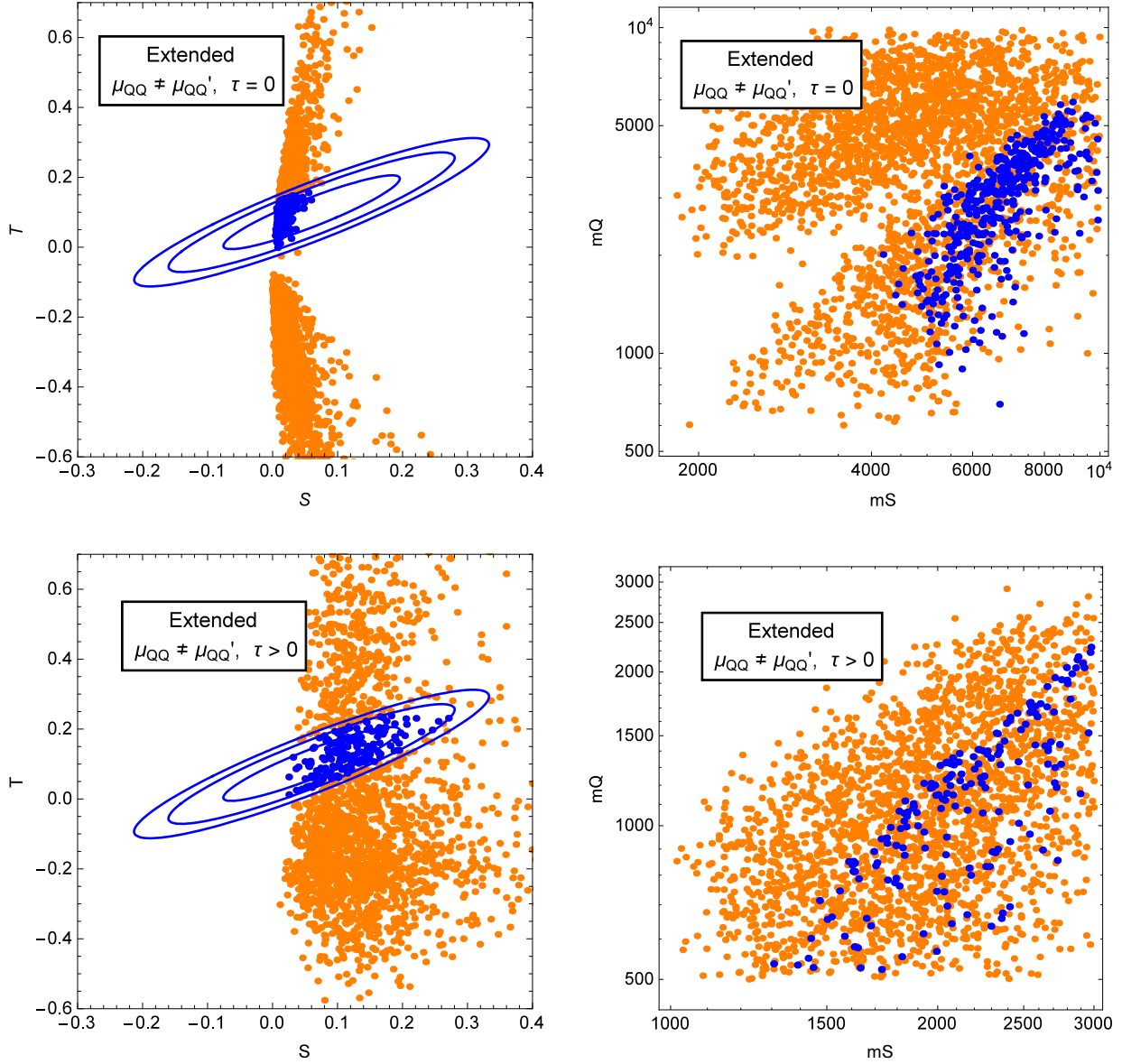

\centering
\psfrag{S}{$S$}
\psfrag{T}{$T$}
\psfragfig[width=0.48\linewidth]{Figures/STextended1}
\hspace{5mm}
\psfrag{x}{$m_S$}
\psfrag{y}{$m_Q$}
\psfragfig[width=0.47\linewidth]{Figures/massesextended} \\ [1.5em]
\psfrag{S}{$S$}
\psfrag{T}{$T$}
\psfragfig[width=0.48\linewidth]{Figures/STextended2}
\hspace{5mm}
\psfrag{x}{$m_S$}
\psfrag{y}{$m_Q$}
\psfragfig[width=0.47\linewidth]{Figures/massesextended2}
\caption{Electroweak precision tests (left) and fermion spectrum (right) for the extended model with $\mu_{QQ}\neq \mu_{QQ}'$. The plots in the upper row assume vanishing tadpole, while those in the lower row have a positive tadpole term. For $\tau > 0$, we also impose $m_Q<m_Q'$. We scan over the ranges $f \in [500,2000]~\rm GeV$, $r_v \in [0.05, 0.95]$, $g_\rho \in [0,3\pi]$, $s_R \in [0,1]$ and $s_L \in [0,1]$. In the plots of the upper row, we fix $m_S$, $m_Q$ and $m'_Q$ from EWSB (Higgs vev and Higgs mass) plus the top mass, but requiring $m_S, m_Q, m'_Q > 500~\rm{GeV}$. In the lower row plots we instead  fix $\tau$, $m_S$ and $m_Q$ from EWSB plus the top mass, requiring $m_S, m_Q > 500~\rm{GeV}$, while scanning over $m'_Q \in [500,3000]~\rm{GeV}$ and fixing $f = 500~\rm{GeV}$. The blue points pass EWPT at 95\% C.L. }
\label{plot:fullmodel}
\end{figure}
\begin{figure}
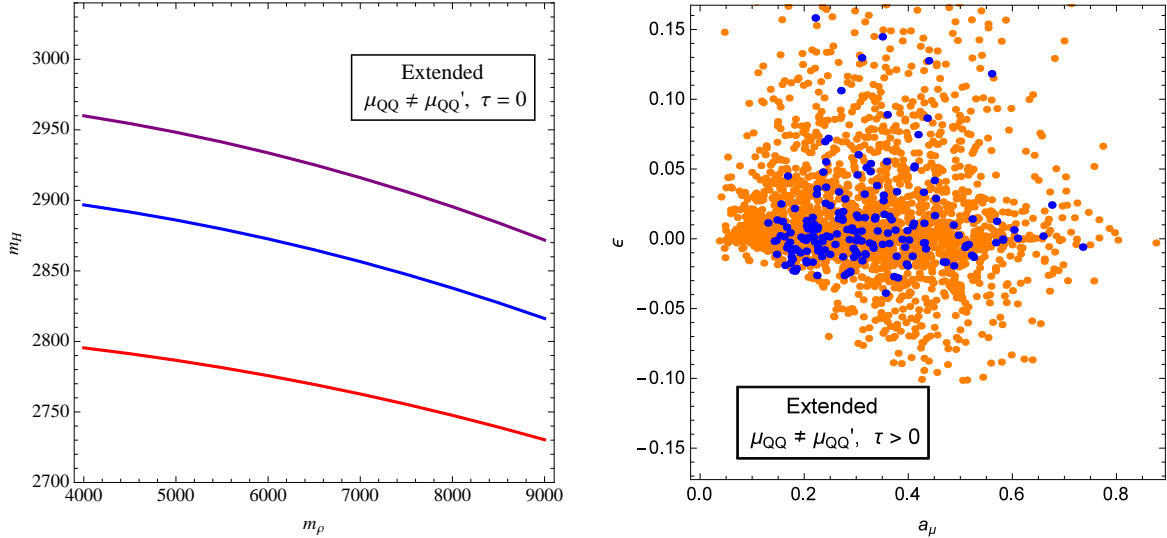

\centering
\psfrag{x}{$m_\rho$}
\psfrag{y}{$m_\H$}
\psfragfig[width=0.45\linewidth]{Figures/prediction}
\hspace{5mm}
\psfragfig[width=0.45\linewidth]{Figures/Gviolation}
\caption{
Left panel: mass of the radial mode (in GeV) as a function of the other masses in the extended model with $\tau=0$. We have fixed $m_Q=1$ TeV, $m_S=5$ TeV and $m_Q'=6.5$ TeV. The ratio $m_\rho/m_a=\sqrt{r_v}$ is held fixed, with $r_v=0.9$, $0.75$ and $0.5$ (from top to bottom). Right panel: amount of explicit violation of the global and custodial symmetries, as parametrized by the quantities $\epsilon$ and $a_\mu$, for the same parameter scan as in the lower row plots of Fig.~\ref{plot:fullmodel}. All points reproduce the correct Higgs, top and $Z$ masses. The blue points pass EWPT at 95\% C.L. }
\label{plot:radialmass}
\end{figure}

The correct Higgs mass and agreement with EWPT can also be achieved in the extended model with $\mu_{QQ}\neq\mu_{QQ}'$ (see Fig.~\ref{plot:fullmodel}). 
As this introduces a new source of explicit $SO(4)$ violation, we expect the $T$ parameter to be affected.
We first consider the case $\tau=0$. 
%
As we already pointed out above, at $\mu_{QQ}=\mu_{QQ}'$, $\beta_{1/2}$ is bounded by Eq.~(\ref{smallbetaextended}), resulting in a too small Higgs mass.
One can show that $\beta$ can be raised if 
\be
(\mu_{QQ}^2-\mu_{QQ}'^2)(\mu_{QQ}^2+\mu_{qQ}^2-\mu_{QQ}'^2)>0\,,
\ee
which implies that  either $\mu_{QQ}>\mu_{QQ}'$ or $m_Q<m_Q'$. 
It turns out that the former case further lowers the fermionic contribution to the $T$ parameter, while the latter one leads to positive $T$.
It is therefore an important conclusion that EWPT force $m_Q<m_Q'$.
We also remark that $m_S$ needs to be comparatively heavy in this scenario. We already pointed out that there exists a lower bound on $m_S$ due to the top mass, $m_S\gtrsim 2f$. The bound is attained at $s_L=1$, $s_R=1/\sqrt 2$ [see Eq.~(\ref{sLsR})]. However at these values, $\beta_{1/2}$ reduces simply to
\be
\frac{\beta_{1/2}}{f^4} =
\frac{3}{4 \pi^2}\frac{m_t^4}{v_{\rm SM}^4}\left(\log\frac{8 f^2}{v_{SM}^2}-\frac{11}{6}\right)~,
\ee
which requires values of $f$ in the multi-TeV range in order to get a large enough Higgs quartic coupling. Larger values of $m_S/f$ are required to avoid this latter conclusion. We typically find that for $f=500$ GeV we need $m_S>2$ TeV, which sets a lower bound on $m_S$ in this scenario. EWPT further increase this bound. We illustrate these conclusions in the upper row plots of Fig.~\ref{plot:fullmodel}.
As we pointed out in Subsection~\ref{sec:pars}, for $\tau=0$ there is one relation between the masses of the model. We show in the left panel of Fig.~\ref{plot:radialmass} the mass of the radial mode as a function of the other masses of the model.

%

Finally, we also performed a full scan of the extended model with $\mu_{QQ}\neq\mu'_{QQ}$ with nonzero tadpole term (see plots in the lower row of Fig.~\ref{plot:fullmodel}). We only discuss in detail the case $\tau>0$. The implications for the spectrum are similar as in the case with $\tau=0$, with the difference that the states can generally be lighter while still passing EWPT. We present in the lower row plots of Fig.~\ref{plot:fullmodel} a scan with fixed $f=500$ GeV, with the fermion masses in the range $\{500,3000\}$ GeV.
In addition we require $m_Q<m_Q'$ as otherwise $T$ is negative. 

It is also interesting to know how much explicit violation of the global symmetry is required in the fermionic mass Lagrangian. We therefore plot in the right panel of Fig.~\ref{plot:radialmass} the quantity $\epsilon$ defined in Eq.~(\ref{mueff2}) against the asymmetry parameter
\be
a_\mu\equiv \frac{\mu'_{QQ}-\mu_{QQ}}{\mu'_{QQ}+\mu_{QQ}}~.
\ee
The point $\epsilon=a_\mu=0$ corresponds to the $SO(5)$ preserving choice $\mu_{QQ}=\mu'_{QQ}=\mu_{tS}$, while the deviations from $a_\mu = 0$ parametrize the breaking of the custodial symmetry in the ``composite sector", to use the language of Section~\ref{top}.
We see that $a_\mu\gtrsim 0.15$ is required in order to obtain points that pass EWPT, while $\epsilon$ is always very small as expected from the general arguments above.

We summarize the various scenarios studied in this section in the following table:
\begin{table}[h]
\centering
\begin{tabular}{ccccccc}
Model&&&$m_h$&EWPT&Spectrum &Remarks\\
\hline
\multirow{2}{*}{Minimal} 	& 		&$\tau=0$		& too light 	&			\\
							& 		&$\tau\neq0$	& \checkmark	& $\times$	\\
\hline
\multirow{6}{*}{Extended} 	& \multirow{3}{*}{$\mu_{QQ}=\mu_{QQ}'$}		& $\tau=0$		& too light 	&			\\
																		&& $\tau>0$	& \checkmark	& 	$\times$	&							& $\epsilon\ll 1$\\
																		&& $\tau<0$	& \checkmark	&	\checkmark	&  $ m_\H<m_S< m'_Q<m_Q$	& $\epsilon\gtrsim 1$\\ \cline{2-7}
							& \multirow{3}{*}{$\mu_{QQ}\neq\mu_{QQ}'$}		& $\tau=0$		& \checkmark 	&	\checkmark		& $m_Q<m_Q',m_S $	& $\epsilon\ll 1$\\
																		&& $\tau>0$	& \checkmark	& 	\checkmark				&$m_Q<m'_Q,m_S$ 	& $\epsilon\ll 1$\\
																		&& $\tau<0$	& \checkmark	&  \checkmark \\
\hline
\end{tabular}
\caption{Summary of our various scenarios. In the last column we have defined $\epsilon=\muext^2/2\mu^2_{tS}$. See text for details.}
\label{tab:models}
\end{table}
%

%
%

\section{Naturalness Considerations}
\label{naturalness}

Indirect constraints from electroweak precision data as well as direct bounds on vectorlike top partners will require a sufficiently high scale for the global symmetry breaking, resulting in a certain fine-tuning of parameters.
In order to get a first idea, it is enough to notice that the largest cancellation occurs in the quantity $\alpha$. There will be a large positive contribution proportional to $\mu_{tS}^2$, leading to a sensitivity
\be
\frac{\Delta\alpha_{\mu_{tS}}}{\alpha}\approx\frac{4r_*\mu_{tS}^2}{r_v m_h^2}~.
\label{est1}
\ee
For $\mu_{tS}= 500$ GeV and $r_v=0.5$ this implies a tuning of about 1\,\%.

 

\begin{figure}
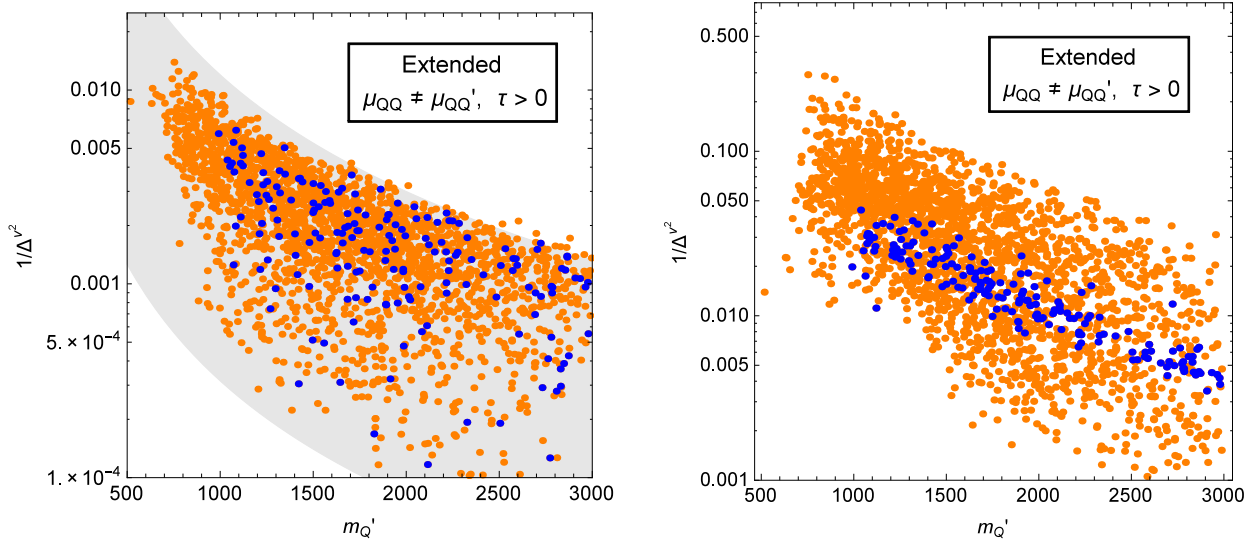

\centering
\psfragfig[width=0.49\linewidth]{Figures/tuning1}
\hspace{5mm}
\psfragfig[width=0.47\linewidth]{Figures/tuning2}
\caption{Left plot: Fine tuning against the mass $m'_Q$ (GeV). The gray band is the estimate Eq.~(\ref{est1}). Right plot: Fine tuning for $\mu_{\rm eff}$ held fixed. 
}
\label{FT}
\end{figure}

In the following we  will quantify these considerations more precisely by evaluating the sensitivity parameter
\be
\Delta^M\equiv\max_P \Delta_P^{M}\,,\qquad \Delta_P^M\equiv\left|\frac{\partial \log M}{\partial\log P}\right|~,
\ee
where $M$ runs over the measured quantities $M\in\{v^2,m_h^2,m_t^2\}$ and $P$ over the parameters of the model. It is important to pick a basis for $P$ that corresponds to the parameters in the Lagrangian. We thus chose 
\be
P\in\{\hat f,f_\rho,\xi, g_\rho ,\mu_{tS},\mu_{QQ},\mu_{QQ'},\tau\}~.
\ee
One can easily evaluate
\be
\Delta^{v^2}_P
=\left|\frac{fs_v}{v\,c_v}\, \partial_{p} \log\frac{\alpha}{\beta}+\partial_p \log f^2\right|\,,\quad
\Delta^{m_h^2}_P=\left|\partial_p \log\frac{\alpha}{f^2}-\frac{s_v^2}{c_v^2}\, \partial_p\log\frac{\alpha}{\beta}\right|\,,\quad
\Delta^{m_t^2}_P=\left|\partial_p \log\frac{\alpha\gamma}{\beta}\right|\,,
\label{Delta}
\ee
where $p=\log P$ and $\gamma=\xi^2\hat f^2\mu_{tS}^2\mu_{qQ}^2/m_S^2m_Q^2$. All of the $\Delta^M$ in Eq.~(\ref{Delta}) are dominated by $\partial_p\log\alpha$, and it turns out the largest one is $\Delta^{v^2}$. We plot the latter in the left panel of Fig.~\ref{FT}, using the same parameter scan as in the lower row plots of Fig.~\ref{plot:fullmodel}. We find that the maximal sensitivity is to the parameter $\mu_{tS}$ for all points, and  pretty much follows the general considerations in Eq.~(\ref{est1}), shown as the gray band in the plot. 
The cancellation of the term proportional to $\mu_{tS}^2$ then typically requires fine tuning below 1\%.

However, as we already discussed in Section~\ref{sec:ewsb}, to some extent this cancellation must happen against the other terms in $\muext$, as the other contributions to $\alpha$ are loop suppressed, and the quantity $\epsilon$ defined in Eq.~(\ref{mueff2}) is expected to be small.
In the vicinity of the point $\mu_{tS}=\mu_{QQ}=\mu_{QQ}'$ this cancellation is protected by the global  symmetry, as except for the mixing term $\mu_{qQ}$ the fermion mass Lagrangian becomes $SO(5)$ symmetric.\footnote{See the right panel of Fig.~\ref{plot:radialmass} for the required amount of $SO(5)$ violation.} However, even away from this point, $\muext$ only renormalizes multiplicatively and one can thus consider this tuning "technically natural". One could thus ask the question of how much tuning is required beyond the one needed in order to obtain small $\epsilon$. This question can be answered by evaluating again the same sensitivity parameters, {\em  but with $\muext$ held fixed}, i.e.~we replace in Eq.~(\ref{Delta})
\be
\partial_p\to \partial_p|_{\muext \rm \ fixed}\,.
\ee 
The result is plotted in the right panel of Fig.~\ref{FT}. 
As expected, the tuning is considerably reduced, as large as 5\% for the points that pass EWPT.
The largest sensitivity occurs with respect to the parameters $g_\rho$ and $\mu_{QQ}'$.  

We now return to theoretical considerations in order to close a possible loophole on how to obtain the set of four-fermion operators used thus far.

\section{A Simple Model leading to $SO(N)$ Symmetric 4-Fermion Interactions}
\label{SO(N)}

We pointed out in Subsection~\ref{NJL} that four-fermion interactions naturally lend themselves to the implementation of $SO(N)$ symmetries, as opposed to $SU(N)$ symmetries. In particular, when $G_S \neq G'_S$ in Eq.~(\ref{L4f}), the global symmetry of the theory is $SO(N)$ and one can work in a region of parameter space where the only light states are those of the $SO(N) \to SO(N-1)$ breaking. In this section we describe a simple renormalizable model that can lead to the above situation. 

Consider a  $SU(N_c) \times SU(N_c)$ gauge theory, and assume that the fermions  described in Subsection~\ref{NJL}, $F^i_L$ and $S_R$, transform in the $(N_c,1)$ representation.  We assume that this gauge symmetry is (spontaneously) broken, as in top-color UV completions~\cite{Hill:1991at} of the top condensation mechanism~\cite{Bardeen:1989ds}. The diagonal $SU(N_c)$, with $N_c = 3$, is then identified with the QCD interactions,\footnote{We imagine that all matter is charged under the first $SU(3)$ group only. Since the full model consists of the SM field content plus additional vectorlike states, the theory is easily seen to be vectorlike with respect to $SU(3) \times SU(3)$, hence anomaly free.} while the broken $SU(N_c)$ induces four-fermion operators below the mass of the corresponding gauge boson, $G_\mu$. In addition, assume there exists a real scalar field $\Xi$ that transforms in the fundamental representation of the ``flavor'' $SO(N)$, as does $F^i_L$, while being a singlet of the new gauge group. In unitary gauge, one can then write the following terms in the UV Lagrangian:
\bea
{\mathcal L}_{\rm UV} &\supset& -\frac{1}{2} M^2_{\Xi} \, \Xi^2 + y \, (\bar{S}_R \, \Xi^i F^i_L+{\rm h.c.}) + \frac{1}{2} \, M^2_G G_\mu G^\mu + \frac{1}{2} \, \hat{g} \, G^A_\mu (\bar{S}_R \gamma^\mu \lambda^A S_R + \bar{F}_{L,i} \gamma^\mu \lambda^A F^i_L)~,
\nn \\
\label{UVModel}
\eea
where $\lambda^A$ are the Gell-Mann matrices. The contractions of the indices not explicitly shown should be obvious. We have assumed above, for simplicity, that the Yukawa coupling, $y$, is real. Integrating out the heavy gauge and scalar fields, leads to an effective Lagrangian
\bea
\mathcal L &\supset& \frac{y^2}{2M^2_{\Xi}} \, (\bar S_R F^i_L+{\rm h.c.})^2 - \frac{\hat{g}^2}{8M^2_G} (\bar{S}_R \gamma^\mu \lambda^A S_R + \bar{F}_{L,i} \gamma^\mu \lambda^A F^i_L)^2
\nn \\
&=&  \frac{y^2}{2M^2_{\Xi}} \, (\bar S_R F^i_L+{\rm h.c.})^2 - \frac{\hat{g}^2}{4M^2_G} \, (\bar{S}_R  \gamma^\mu \lambda^A S_R) (\bar{F}_{L,i} \gamma^\mu \lambda^A F^i_L) + \cdots
~,
\label{CrossedChannels}
\eea
where the terms not shown correspond to vector channels and are further discussed in App.~\ref{app:SO(N)}.
Upon Fierz rearrangement of the second term, and neglecting further vector channels, one gets
\bea
\mathcal L &\supset&  \frac{y^2}{2M^2_{\Xi}} \, (\bar S_R F^i_L+{\rm h.c.})^2 + \frac{\hat{g}^2}{M^2_G} \, (\bar{S}_R  F^i_L) (\bar{F}_{L,i} S_R) + \cdots~.
\label{scalarchannel}
\eea
This can be written as ${\cal L}_S + {\cal L}'_S$ of Eq.~(\ref{L4f}) with the identifications
%
\bea
G_S &=& \frac{\hat{g}^2}{2M^2_G} +  \frac{y^2}{M^2_{\Xi}}~,
\hspace{1cm}
G'_S ~=~ \frac{\hat{g}^2}{2M^2_G}~.
\eea
We see that the splitting between $G_S$ and $G'_S$ [hence the explicit breaking from $SU(N)$ down to $SO(N)$] is controlled by the heavy scalar sector [which is not surprising given that $\Xi$ is the only field that transforms explicitly only under $SO(N)$]. In addition, we see that the effective coupling $G_S$ is positive and naturally larger than $G'_S$, as required in Subsection~\ref{NJL}.  We conclude that it is not far-fetched to have $G_S$ close to criticality (through some degree of tuning, as is typical in the context of the NJL mechanism), while $G'_S$ is sub-critical so that the corresponding (irrelevant) four-fermion operators play no role at low energies.

One should note that the scalar field $\Xi$ enters in a very similar fashion to the scalar $\Phi$ of Subsection~\ref{NJL}. However, these d.o.f.~are distinct and should not be confused. Whereas the scalar $\Xi$ above is a propagating degree of freedom, with a well-defined kinetic term above scales of order $G_S^{-1/2}$, the scalar $\Phi$ discussed in the main text is a $(\bar{F}_L S_R)$ bound state, that exists as such only well below the scale $G_S^{-1/2}$. While the two scalars share the same quantum numbers, they are better thought as ``dual'' to each other, in the sense of UV versus IR degrees of freedom. Having said this, it is possible that the scalar $\Xi$ is itself a composite that arises in the process of breaking the $SU(N_c) \times SU(N_c)$ gauge symmetry and whose mass is tied to that of the gauge field $G_\mu$. Since for our applications $\Xi$ can have a mass of a similar order as that for $G_\mu$ (but with no special relation), we see that no tuning beyond the one required to lie close to criticality needs to be imposed in this new scalar sector. 


\section{Brief Phenomenological Remarks}
\label{pheno}

We end by offering a few comments on the expectations for present and future colliders within our scenario. A more detailed study will be presented elsewhere~\cite{Gersdorff15}. We start by noting that, although we are providing a description of the Higgs constituents and the interactions that bind them together (the NJL model discussed in most of this paper), the low-energy physics is expected to be well-described by a general non-linear $\sigma$-model based on the $SO(5)/SO(4)$ symmetry breaking pattern. Uncovering the specific manifestations of our set-up requires experiments that are sensitive to higher energies. 

The model we have described contains a rather minimal set of fields beyond those of the SM. In the fermion sector, we have a vectorlike $5$ of $SO(5)$, which decomposes into a bi-doublet of $SO(4) \approx SU(2)_L \times SU(2)_R$ [the states $\Q$ and $\Q'$ with approximate masses $m_Q$ and $m'_Q$, as given in Eqs.~(\ref{massesf}) and (\ref{massesf2})] plus a singlet $S$ with approximate mass as given in Eq.~(\ref{massesf}). Hence, we have three top-like partners, a $b'$ and an exotic $Q = 5/3$ state, as is common in many CH model with a custodial protection of $Zb\bar b$. Several search strategies to look for such states at the LHC have been put forward (see, for instance,~\cite{Gripaios:2014pqa}, and references therein), and a number of LHC searches already exist which can be sensitive to such states below about $800~{\rm GeV}$~\cite{Chatrchyan:2013wfa,Chatrchyan:2013uxa,ATLASTopPartner8TeVCONF,CMSBSearch}. We have found that typically $m_Q < m_{Q}' , m_S$.\footnote{Although in the presence of a very large tadpole, Eq.~(\ref{tadpole}), one can have the peculiar hierarchy $m_S < m'_Q < m_Q$. In this case, all states would be rather heavy.} Direct searches, in particular those for $Q=-1/3$ and $Q = 2/3$ states (in our case arising from the $\Q$ state), imply within our model that $m_S \gtrsim 1.6~{\rm TeV}$ (although it may well be significantly heavier), while EWPT require $m_Q' \gtrsim 1~{\rm TeV}$ (which corresponds to the mass of the $Q=5/3$ state). 

In addition to the fermion sector above, we predict the existence of a ``radial" Higgs mode, $\H$, that together with the pNGB degrees of freedom form a $5$ of $SO(5)$.
 Such a scalar state would likely be beyond the reach of the LHC, but might be accessible at a 100~TeV collider.  If $\H$ happens to be a sufficiently narrow resonance, one could try to measure its Yukawa coupling $\xi$ with the previously mentioned fermionic resonances. One then expects this coupling to become large as one approaches a certain scale $\Lambda$. 
We also reiterate that the mass of the radial mode is determined by the Yukawa coupling $\xi$ and the symmetry breaking scale $f$. This is a prediction of the NJL model, in sharp contrast to an arbitrary linearization of the sigma model (as in the SM).

Finally, there are spin-1 resonances that, however, cannot be guaranteed to lie below the cutoff of the theory, where the composite Higgs ``dissolves" into its constituents. The gap between this cutoff scale and $m_\H$ depends on how close to criticality is the (scalar channel) four-fermion coupling. Near this cutoff scale, there may exist further excited resonances of various spins. 

The fact that in our case the Higgs is a bound state of fermions
closely connected to the top quark (i.e.~some of the resonances mentioned above) should have, in principle, measurable consequences below the scale $\Lambda$ (where $\xi$ blows up). This, together with evidence for 4-fermion interactions involving these 
fermionic resonances, with a strength $G_S$ of order $1/\Lambda^2$, would further point to a picture as studied in this paper.

We have already noted that above this scale the UV degrees of freedom may be relatively few, and described by a renormalizable theory, perhaps valid up to much higher energies.
If the model of section 6 was indeed the underlying microscopic physics giving rise to the 4-fermion interactions, one could expect those states to be not too far above  the scale of strong coupling indicated by the Yukawa interaction $\xi$. Carrying out such a program would require energies beyond the LHC, and detailed studies are necessary before one can judge its feasibility, and to what extent (and with what type of machines) one could establish such a picture. Here we only point out a few avenues that need to be explored in order to eventually test the model. 

In terms of Higgs physics, at higher-energies one would expect to start seeing form factors that indicate its composite nature. Perhaps in the details of such form factors there could be a measurable  imprint of the NJL dynamics. Studying such an interesting possibility goes beyond the scope of this paper, which was simply to provide a step towards establishing microscopic realizations of modern pNGB scenarios. We leave a detailed phenomenological study for future work.

\section{Conclusions}
\label{conclusions}

One of the major questions whose answer will determine qualitatively the nature of physics near the weak scale is whether the Higgs resonance identified by the ATLAS and CMS collaborations~\cite{Aad:2012tfa,Chatrchyan:2012ufa} is elementary down to ultra-short distances or whether it is actually a composite state of more fundamental constituents, whose nature could be revealed at energies not far above the EW scale. In either case, a phenomenon never seen in nature before would have been established. In the first case, we would have discovered the first example of an ``elementary'' scalar, which perhaps could suggest the presence of supersymmetry at some higher scale (given that fundamental scalars are a generic prediction of supersymmetric scenarios). If, on the other hand, the Higgs turned out to be composite, it would be the first example of condensation by such a scalar that can be effectively described by weakly coupled dynamics. In spite of this, its composite nature itself would point to some strongly coupled underlying dynamics that, if realized by nature, would lead to a qualitatively different view of the EW scale than in the first case. 

In this work we have explored a possible microscopic realization of the second possibility, modeling the interactions that lead to the scalar bound state via four-fermion interactions \`a la NJL~\cite{Nambu:1961tp,Nambu:1961fr}.  This allows the explicit identification of the Higgs constituents and the interactions responsible for the formation of the bound states. As was first done in Ref.~\cite{Cheng:2013qwa}, in order to obtain a scalar resonance that is parametrically lighter than other strong resonances, it is assumed that the microscopic Lagrangian possesses an approximate global symmetry that is spontaneously broken by the NJL mechanism, thus leading to suitable Nambu-Goldstone bosons, some of which can be identified with the Higgs. We are therefore providing a possible UV completion to a class of pNGB Higgs scenarios that have been studied only at the level of the non-linear $\sigma$-model.  In order to control potentially large corrections to precision EW measurements, we implement a scheme that preserves approximately a custodial symmetry, as well as the custodial symmetry protection of the $Zb\bar{b}$ coupling put forward in~\cite{Agashe:2006at} (a similar setup to ours was studied in~\cite{Cheng:2014dwa}). As  is well known, minimality together with the above requirements uniquely singles out the symmetry breaking pattern $SO(5) \to SO(4) \approx SU(2)_L \times SU(2)_R$. In addition, the SM $b_L$ satisfies $T^3_L = T^3_R$.

We point out here that four-fermion interactions naturally lend themselves to implementing the above symmetry breaking pattern. Unlike fermion bilinears\footnote{Kinetic terms and gauge interactions. Mass terms can in principle reduce the symmetry appropriately. However, in our minimal implementations this is not an option.} which typically preserve an $SU(N)$ symmetry, for four-fermion operators it is possible to impose a reality condition that preserves only the $SO(N)$ subgroup. We also point out that in the context of the NJL mechanism, the breaking by the 4-fermion interactions from $SU(N)$ down  to $SO(N)$ may be small, yet effective: if one is close to criticality, small effects can easily make the additional $SU(N)$ resonances much heavier than those associated with the $SO(N)$ breaking. In this respect, our implementation represents a considerable simplification with respect to the schemes proposed in~\cite{Cheng:2013qwa,Cheng:2014dwa}. A possible question is how to generate the four-fermion interactions with the required global $SO(N)$ symmetry. As an existence proof, we have provided a simple renormalizable model  that generates such 4-fermion interactions, thus realizing our framework.

Also, unlike the previous works, we have  explicitly described the spin-1 sector that would be responsible for the cancellation of the quadratic divergences in the Higgs mass arising from the SM gauge interactions. We have highlighted the qualitative differences between the spin-1 and scalar sectors in terms of calculability. We have paid particular attention to the fermion sector with an eye towards minimality, emphasizing the role played by the new degrees of freedom beyond the SM. Importantly, we find that the presence of quasi-fixed points (generalizing the observations in~\cite{Bardeen:1989ds}) imply that the low-energy physics is largely insensitive to uncertainties arising from the underlying strong dynamics.

Given the presence of four-fermion interactions that, through the NJL mechanism, break $SO(5) \to SO(4)$, we recover at low energies the SM field content, except that the Higgs potential is dynamically induced and the breaking of the EW symmetry (or not) is an outcome of the theory. We have shown that in regions of parameter space where the EW symmetry is actually broken, it is possible to accommodate the observed Higgs mass of about $125$~GeV~\cite{Aad:2014aba,CMS:2014ega}. In some cases, this requires turning on a tadpole term (equivalent to a certain vector-like fermion mass), but we have also identified cases where such a tadpole is not essential. We find, however, that in general there is some tension with EW precision measurements. This is due to a typically negative contribution to the Peskin-Takeuchi~\cite{Peskin:1991sw} $T$-parameter that arises from two sources: the non-linear couplings of the Higgs field to the gauge bosons, due to its pNGB nature~\cite{Barbieri:2007bh}, and the fermion loop contributions. The latter effect is tied to the imposed custodial protection (on $T$ and $Zb\bar{b}$) which suppresses the contributions to an acceptable order of magnitude, but typically leads to a negative sign that makes agreement with the $S-T$ ellipse challenging. Such an effect has the same origin as first pointed out in the context of extra-dimensional constructions in~\cite{Carena:2006bn}. Interestingly, in our setup it is possible to allow for a controlled amount of custodial breaking in the heavy fermion sector so that the corresponding contribution to $T$ changes sign and allows compatibility with EWPT. We emphasize that this custodial breaking is soft, so that the effects are fully calculable within the theory, and no particular tuning of parameters is required. Nevertheless, well-defined regions of parameter space are selected in this way, which have implications for the expected hierarchies of states beyond the SM ones.

We have also studied the tuning in the present models, as measured by the sensitivity of low-energy observables to UV Lagrangian parameters. Without any special assumptions about the UV, the tuning so defined is found to be at the percent level or worse. Interestingly, however, this tuning has a rather well-defined source throughout the region of parameter space. It arises from a necessary cancellation between fermion mass parameters, forced by current LHC direct lower bounds on some of these states. However, the region corresponding to such a cancellation actually corresponds to an enhanced symmetry point of the theory. Therefore, small deviations from perfect cancellation can be naturally small, in the 't Hooft sense. The remnant fine-tuning (assuming the previous cancellation between fermion mass parameters) can be on the order of $1-5\%$.

We also presented a simple model for the origin of the new four-fermion interactions with the correct global symmetry breaking pattern.
Possible phenomenological consequences of our setup, such as the existence of the radial mode $\H$, of vector resonances, and of the various fermionic resonances, as well as possible changes in the Higgs couplings due to form factors, were briefly mentioned.

We end by noting that we have focused our attention on the third generation, which is likely to be most relevant for EWSB. However, accomodating the first two generations, and understanding how issues such as the flavor structure could be embedded in the present framework would be of extreme interest, and are likely to provide further handles from precision flavor measurements, rare decays or CP-violation.

\acknowledgments
This work was supported by the S\~ao Paulo Research Foundation (FAPESP)
under grant \#2011/11973. G.G.~was supported by FAPESP grant \# 2012/22639.
R.R.~was partially supported by a CNPq research grant.
\appendix

\section{Four-fermion Interactions from Vector Boson Exchange}
\label{app:SO(N)}

In this appendix we rewrite the four-fermion interactions induced by heavy vector boson exchange in the model of Section~\ref{SO(N)}. Our purpose is to identify the full set of scalar and vector channels thus generated (the scalar ones were presented in the main text). For this purpose, it will be useful to recall the following generic Fierz rearrangement for the vector channel 
\bea
(\bar A\gamma^\mu B )( \bar C\gamma_\mu D) &=&-(\bar A D)(\bar CB )
+(\bar A \gamma^5 D)(\bar C\gamma^5 B )
\nn \\ [0.5em]
& & \mbox{} +\frac{1}{2} \left[
(\bar A \gamma^\mu D)(\bar C\gamma_\mu B )
+(\bar A \gamma^5\gamma^\mu D)(\bar C\gamma^5\gamma_\mu B )
\right]~,
\eea
as well as the ``color'' identity
\be
(\bar \chi_i \Gamma \chi^j)(\bar \psi_j \Gamma \psi^i)=(\bar\chi \, \Gamma \, t^A\chi)(\bar\psi \, \Gamma \, t^A\psi)+\frac{1}{N}  \, (\bar\chi \Gamma \chi)(\bar\psi \Gamma \psi)~,
\ee
where $\Gamma$ is any set of gamma matrices that makes $M_i^{\,\,\, j} = (\bar A_i \Gamma A^j)$ hermitian, while $i,j$ are $U(N)$ indices.
Here and  in the following we normalize the $SU(N)$ generators as $\tr t^A t^B=\delta^{AB}$.

In the following, we denote by $i$ ($I$) the fundamental (adjoint) index of $SU(N_L)$, and by $a$ ($A$) the fundamental (adjoint) index of $SU(N_c)$. We focus on the fermion field content of Section~\ref{SO(N)}, $F_L^{a,i}$ and $S_R^a$, so that, for example,
\bea
\frac{1}{2} \, (\bar F_L \gamma^\mu  \lambda^A F_L)^2 &=& (\bar F_{L,a} \gamma^\mu F^b_L)^2-\frac{1}{N_c}( \bar F_L \gamma^\mu F_L)^2
\nn \\
&=&(\bar F_{L,i} \gamma^\mu F^j_L)^2-\frac{1}{N_c}(\bar F_L \gamma^\mu F_L)^2
\nn \\
&=&(\bar F_L \gamma^\mu T^I F_L)^2+\left( \frac{1}{N_L}-\frac{1}{N_c} \right)(\bar F_L \gamma^\mu F_L)^2~,
\eea
where $T^I$ are the generators of $SU(N_L)$. Here we have used a condensed notation where the color and flavor indices in $F_L^{a,i}$ that are not shown explicitly are understood to be appropriately  contracted between the fields within each factor in parenthesis.

Analogously for the RH field $S^a_{R}$ we find
\bea
\frac{1}{2} \, (\bar S_R \gamma^\mu \lambda^A S_R)^2 &=& \left(1-\frac{1}{N_c} \right)(\bar S_R \gamma^\mu S_R)^2~,
\eea
while for the mixed term one obtains
\bea
\frac{1}{2} \, (\bar S_R \gamma^\mu \lambda^A S_R) (\bar F_L \gamma^\mu \lambda^A F_L)
&=&(\bar S_{R,a}\gamma^\mu S^b_R) (\bar F_{L,b}\gamma_\mu F^a_L) 
-\frac{1}{N_c}(\bar S_R \gamma^\mu S_R)(\bar F_L \gamma_\mu F_L )
\nn \\
&=&-2(\bar S_R F^i_L)(\bar F_{L,i} S_R)
-\frac{1}{N_c}(\bar S_R \gamma^\mu S_R)(\bar F_L \gamma_\mu F_L)~,
\eea
where the first term reproduces the scalar channel written in Eq.~(\ref{scalarchannel}).
Putting everything together one concludes that the exchange of the heavy vector bosons $G_\mu$ induces the following set of four-fermion interactions: 
\bea
\frac{1}{2} \, (\bar S_R \gamma_\mu\lambda^A S_R+\bar F_L \gamma_\mu \lambda^A F_L)^2 &=&
-4(\bar S_R F^i_L)(\bar F_{L,i} S_R)+(\bar F_L \gamma_\mu T^I F_L )^2
\nn \\ [0.2em]
& & -\frac{1}{N_c}(\bar S_R \gamma_\mu S_R +\bar F_{L} \gamma_\mu F_L)^2
+\frac{1}{N_L}(\bar F_{L} \gamma_\mu F_L)^2+(\bar S_R\gamma_\mu S_R)^2
\nn \\ [0.6em]
&=&
-4(\bar S_R F^i_L)(\bar F_{L,i} S_R)+(\bar F_L \gamma_\mu T^I F_L )^2
\label{vectorFierz}
 \\ [0.5em]
&+& \left(\frac{1}{4N_L}+\frac{1}{4}-\frac{1}{N_c} \right) J_+^2
+ \left(\frac{1}{4N_L}+\frac{1}{4} \right)J_-^2+
\left(\frac{1}{2N_L}-\frac{1}{2} \right)J_+J_-
\nn
\eea
where we defined
\be
J_{\pm\,\mu}\equiv (\bar F_L \gamma_\mu F_L \pm \bar S_R \gamma_\mu S_R)~.
\ee
We see that apart from the scalar channel discussed in Section~\ref{SO(N)}, the topcolor interactions also produce a vector $SU(N_L)$ channel that contains the $SO(N)_L$ channel of Eq.~(\ref{currents}) [with the correct sign required in the discussion around~Eq.(\ref{currents})].
There are also $U(1)$ currents that, however, do not exactly match the $U(1)_X$ current of Eq.~(\ref{currents2}). Had we included $N_R$ flavor $S_R^{a,i}$, the last line in Eq.~(\ref{vectorFierz}) would have read
\bea
\left(\frac{1}{4N_L}+\frac{1}{4N_R}-\frac{1}{N_c} \right) J_+^2
+ \left(\frac{1}{4N_L}+\frac{1}{4N_R} \right)J_-^2+
\left(\frac{1}{2N_L}-\frac{1}{2N_R} \right)J_+J_-~.
\nn
\eea
The condition for the eigenvalues of this system to be both positive is $N_c>N_L+N_R$. The condition for $J_+$ and $J_-$ to be eigenstates is $N_L=N_R$. For $N_c = 3$ and the flavor content we have in mind, there is always one negative and one positive eigenvalue. The ``wrong sign" four-fermion interaction would then correspond to a repulsive channel that does not lead to spin-1 bound states. At any rate, as we remarked in the main text, the $U(1)_X$ resonance leads only to subleading modifications when $g^2_X \gg g_0'^2$. Therefore, for our purposes the microscopic model introduced in Section~\ref{SO(N)} is sufficient to establish the possibility of UV completing our point of departure in the main text.

\section{$SO(5)$ Basis}
\label{GroupTheory}

The Minimal Composite Higgs literature based on the coset space $SO(5) / SO(4)$ has widely adopted the basis introduced in~\cite{Agashe:2004rs}, given by:
\bea
(T^a_{L,R})_{ij} &=& -\frac{i}{2} \left[ \frac{1}{2} \epsilon^{abc} \left( \delta^b_i \delta^c_j - \delta^b_j \delta^c_i \right) \pm \left( \delta^a_i \delta^4_j - \delta^a_j \delta^4_i \right) \right]~,
\nonumber \\
T^{\hat{a}}_{ij} &=& - \frac{i}{\sqrt{2}} \left( \delta^{\hat{a}}_i \delta^5_j - \delta^{\hat{a}}_j \delta^5_i \right)~.
\label{StandardBasis}
\eea
However, we find it more convenient to use the basis obtained by a similarity transformation defined by the unitary matrix:
\bea
P &=& 
\frac{1}{\sqrt{2}} 
\left(
\begin{array}{ccccc}
0  & 0  & i  & 1  & 0  \\
i  & -1  & 0  & 0  & 0  \\
i  & 1  & 0  & 0  & 0  \\
0  & 0  & -i  & 1  & 0  \\
0  & 0  & 0  & 0  & \sqrt{2}  
\end{array}
\right)~.
\eea
Explicitly we have:
\begin{eqnarray}
T^{1}_L=\left(
\begin{array}{ccccc}
 0 & \frac{1}{2} & 0 & 0 & 0 \\
 \frac{1}{2} & 0 & 0 & 0 & 0 \\
 0 & 0 & 0 & \frac{1}{2} & 0 \\
 0 & 0 & \frac{1}{2} & 0 & 0 \\
 0 & 0 & 0 & 0 & 0
\end{array}
\right)~, 
\ 
T^{2}_L=\left(
\begin{array}{ccccc}
 0 & -\frac{i}{2} & 0 & 0 & 0 \\
 \frac{i}{2} & 0 & 0 & 0 & 0 \\
 0 & 0 & 0 & -\frac{i}{2} & 0 \\
 0 & 0 & \frac{i}{2} & 0 & 0 \\
 0 & 0 & 0 & 0 & 0
\end{array}
\right)~, 
\ 
T^{3}_L=\left(
\begin{array}{ccccc}
 \frac{1}{2} & 0 & 0 & 0 & 0 \\
 0 & -\frac{1}{2} & 0 & 0 & 0 \\
 0 & 0 & \frac{1}{2} & 0 & 0 \\
 0 & 0 & 0 & -\frac{1}{2} & 0 \\
 0 & 0 & 0 & 0 & 0
\end{array}
\right)\ , \nonumber
\end{eqnarray}
\begin{eqnarray}
T^{1}_R= - \left(
\begin{array}{ccccc}
 0 & 0 & \frac{1}{2} & 0 & 0 \\
 0 & 0 & 0 & \frac{1}{2} & 0 \\
 \frac{1}{2} & 0 & 0 & 0 & 0 \\
 0 & \frac{1}{2} & 0 & 0 & 0 \\
 0 & 0 & 0 & 0 & 0
\end{array}
\right)~, 
\ 
T^{2}_R=\left(
\begin{array}{ccccc}
 0 & 0 & -\frac{i}{2} & 0 & 0 \\
 0 & 0 & 0 & -\frac{i}{2} & 0 \\
 \frac{i}{2} & 0 & 0 & 0 & 0 \\
 0 & \frac{i}{2} & 0 & 0 & 0 \\
 0 & 0 & 0 & 0 & 0
\end{array}
\right)~, 
\ 
T^{3}_R=\left(
\begin{array}{ccccc}
- \frac{1}{2} & 0 & 0 & 0 & 0 \\
 0 & -\frac{1}{2} & 0 & 0 & 0 \\
 0 & 0 & \frac{1}{2} & 0 & 0 \\
 0 & 0 & 0 & \frac{1}{2} & 0 \\
 0 & 0 & 0 & 0 & 0
\end{array}
\right)\ , \nonumber
\end{eqnarray}
\begin{eqnarray}
T^{\hat 1}=\left(
\begin{array}{ccccc}
 0 & 0 & 0 & 0 & 0 \\
 0 & 0 & 0 & 0 & \frac{1}{2}  \\
 0 & 0 & 0 & 0 & \frac{1}{2}  \\
 0 & 0 & 0 & 0 & 0 \\
 0 & \frac{1}{2}  & \frac{1}{2}  & 0 & 0
\end{array}
\right)~, 
\ 
T^{\hat 2}=\left(
\begin{array}{ccccc}
 0 & 0 & 0 & 0 & 0 \\
 0 & 0 & 0 & 0 & \frac{i}{2}  \\
 0 & 0 & 0 & 0 & -\frac{i}{2}  \\
 0 & 0 & 0 & 0 & 0 \\
 0 & -\frac{i}{2}  & \frac{i}{2}  & 0 & 0
\end{array}
\right)~,
\nonumber
\end{eqnarray}
\begin{eqnarray}
T^{\hat 3}=\left(
\begin{array}{ccccc}
 0 & 0 & 0 & 0 & \frac{1}{2} \\
 0 & 0 & 0 & 0 & 0  \\
 0 & 0 & 0 & 0 & 0  \\
 0 & 0 & 0 & 0 & -\frac{1}{2} \\
 \frac{1}{2} & 0 & 0  & -\frac{1}{2} & 0
\end{array}
\right)~, 
\ 
T^{\hat 4}=\left(
\begin{array}{ccccc}
 0 & 0 & 0 & 0 & -\frac{i}{2} \\
 0 & 0 & 0 & 0 & 0  \\
 0 & 0 & 0 & 0 & 0  \\
 0 & 0 & 0 & 0 & -\frac{i}{2} \\
 \frac{i}{2} & 0 & 0  & \frac{i}{2} & 0
\end{array}
\right)~.
\label{OurBasis}
\end{eqnarray}
Then, for instance, the fermionic $5_{2/3}$ containing the LH top and bottom reads:
\bea
\frac{1}{\sqrt{2}}
\left(
\begin{array}{ccc}
-i (b_L + \chi_L)   \\
- b_L + \chi_L   \\
-i(t_L - t'_L)   \\
t_L + t'_L  \\
\sqrt{2} \, S_L
\end{array}
\right)
&\stackrel{P}{\longrightarrow}&
\left(
\begin{array}{ccc}
t_L   \\
b_L   \\
\chi_L   \\
t'_L  \\
S_L
\end{array}
\right)
\hspace{5mm}
~\Longleftrightarrow~
\left(
\begin{array}{ccc}
t_L  &  \chi_L   \\
b_L  & t'_L  
\end{array}
\right)
~\oplus~S_L~,
\label{Fermionic5}
\eea
where the first vector is in the basis of Eq.~(\ref{StandardBasis}), the second in our basis, Eqs.~(\ref{OurBasis}), and the last column makes the connection to the $SO(4) \simeq SU(2)_L \times SU(2)_R \subset SO(5)$ notation. In the main text, we have used the notation $Q^1_L = (t_L, b_L)^T$, $Q^2_L = (\chi_L, t'_L)^T$.

For the Higgs field, we have:
\bea
\left(
\begin{array}{ccc}
h_1   \\
h_2   \\
h_3   \\
h_4  \\
\hat{f}
\end{array}
\right)
&\stackrel{P}{\longrightarrow}&
\frac{1}{\sqrt{2}}
\left(
\begin{array}{ccc}
h_4 + i h_3   \\
-h_2 + i h_1   \\
h_2 + i h_1   \\
h_4 - i h_3  \\
\sqrt{2} \, \hat{f}
\end{array}
\right)
~\equiv~
\left(
\begin{array}{ccc}
H^{0*}   \\
-H^-   \\
H^+   \\
H^0  \\
\hat{f}
\end{array}
\right)
\hspace{5mm}
~\Longleftrightarrow~
\left(
\begin{array}{ccc}
H^{0*}  &  H^+   \\
-H^-  & H^0  
\end{array}
\right)
\eea
In the main text, we have used the notation $\tilde{\phi} = (H^{0*}, -H^-)^T$, $\phi = (H^+, H^0)^T$.

\section{Computation of the pNGB Potential}
\label{app:potential}

In this appendix we collect a few details of the computation of the pNGB Higgs effective potential, focusing on the contributions to the coefficients $\alpha$ and $\beta$ defined in Eq.~(\ref{alphabeta}).
These are obtained by evaluating the pNGB effective potential in the (unitary gauge) constant backgrounds, $U_{\bf 1} = 1$ and [see Eq~(\ref{OurBasis})]
\bea
U_{\bf 5} &=& e^{\sqrt{2} i \, h T^{\hat{4}} / f}
~=~
\left(
\begin{array}{ccccc}
\cos^2\left(\frac{h}{2 f}\right) & 0 & 0 & -\sin ^2\left(\frac{h}{2 f}\right) & \frac{\sin \left(\frac{h}{f}\right)}{\sqrt{2}} \\
 0 & 1 & 0 & 0 & 0 \\
 0 & 0 & 1 & 0 & 0 \\
-\sin ^2\left(\frac{h}{2 f}\right) & 0 & 0 & \cos ^2\left(\frac{h}{2 f}\right) & \frac{\sin \left(\frac{h}{f}\right)}{\sqrt{2}} \\
-\frac{\sin \left(\frac{h}{f}\right)}{\sqrt{2}} & 0 & 0 & -\frac{\sin \left(\frac{h}{f}\right)}{\sqrt{2}} & \cos \left(\frac{h}{f}\right) \\
\end{array}
\right)~,
\label{shDef}
\eea
after expanding for small  $s_h = \sin(h/f)$.  We treat separately the contributions from the states of spin-1, 1/2 and 0.

\subsection{Vector Resonances}
\label{app:vector}

As is commonly done in CH models, we integrate over the vector resonances. The elementary gauge fields will pick up form factors. In momentum space, the relevant spin-1 Lagrangian reads (see, for instance, \cite{Marzocca:2012zn})
\be
\mathcal L_G~=~\frac{P^{\mu\nu}}{2} \left(\Pi_W W_\mu^a W_\nu^a
 +\Pi_BB_\mu B_\nu
+\Pi_1\frac{s_h^2}{4}
[W_\mu^a W_\nu^a+B_\mu B_\nu-B_\mu W_\nu^3-W_\nu^3 B_\mu]
\right)~,
\ee
with $P_{\mu\nu}=\eta_{\mu\nu}-p_\mu p_\nu/p^2$ and
\bea
\Pi_W(p^2) &=& -\frac{p^2}{g_0^2}+\frac{p^2}{g_\rho^2}\, \frac{m_\rho^2}{p^2-m_\rho^2}~,\qquad
\Pi_B(p^2)~=~-\frac{p^2}{g_0'^2}+\frac{p^2}{g_\rho^2}\, \frac{m_\rho^2}{p^2-m_\rho^2}
+\frac{p^2}{g_X^2}\, \frac{m_X^2}{p^2-m_X^2}~,
\nn \\ [0.4em]
\Pi_1(p^2) &= &f^2+p^2\left[\frac{f_\rho^2-f^2}{p^2-m_a^2}-\frac{f_\rho^2}{p^2-m_\rho^2}\right]
~=~\frac{f^2m_a^2m_\rho^2}{(p^2-m_\rho^2)(p^2-m_a^2)}~,
\label{spin1formfactor}
\eea
where Eq.~(\ref{vectormasses}) has been used.
The NGB potential can then be written as
\be
V_{V}~=~\frac{3}{32 \pi^2}\int dp^2\, p^2\left[2\log\left(1+\frac{s_h^2}{4}\frac{\Pi_1(-p^2)}{\Pi_W(-p^2)}\right)+\log\left(1+\frac{s_h^2}{4}\left[\frac{\Pi_1(-p^2)}{\Pi_W(-p^2)}+\frac{\Pi_1(-p^2)}{\Pi_B(-p^2)}\right]\right)\right]
\ee
corresponding to the contributions of the $W$ and $Z$ bosons respectively. It is manifestly UV finite.

In the parametrization of Eq.~(\ref{alphabeta}) one has
\bea
\alpha_1 &=& -\frac{3}{64 \pi^2}f^2m_\rho^2\left(3g^2
+g'^2
\right)\frac{\log r_v}{r_v-1}~,
\nn\\ [0.4em]
\beta_1 &=&-\frac{3f^4}{16(4 \pi)^2}\left(
2g^4\log\frac{m_\rho^2}{m_W^2(h)}
+(g^2+g'^2)^2\log\frac{m_\rho^2}{m_Z^2(h)}
\right.
\nn\\ [0.3em]
&&
\left.
+\left[2g^4+(g^2+g'^2)^2\right]
\left[\frac{(3-r_v)r_v^2\log r_v}{(r_v-1)^3}-\frac{r_v^2+1}{(r_v-1)^2}\right]
\right)~,
\eea
where we have approximated $g_\rho^2, g_X^2 \gg g_0^2\,, g_0'^2$, i.e.~$g^2\approx g_0^2$ and $g'^2\approx g_0'^2$ and $r_v=m_\rho^2/m_a^2<1$. Note that the $U(1)_X$ sector does not enter at leading order in these expansions.

\subsection{Fermion Resonances}
\label{app:fermion}

The fermion mass matrix is most easily evaluated in the unrotated (non-HLS) basis, where
\be
\tilde \phi~=~ \frac{1}{\sqrt{2}} \begin{pmatrix} s_h \\0 \end{pmatrix} \hat f~, \qquad
\phi~=~ \frac{1}{\sqrt{2}} \begin{pmatrix}0\\ s_h\end{pmatrix}\hat f~,\qquad
\phi_5~=~c_h\hat f~.
\ee
Substituting these fields, the mass terms are given as
\be
\mathcal L_{-1/3}=-\mu_{QQ}\, \bar Q_L^{1,2}Q_R^{1,2} + {\rm h.c.}\,,\qquad \mathcal L_{5/3}=-\mu_{QQ}'\, \bar Q_L^{2,1}Q_R^{2,1} + {\rm h.c.}~,
\ee
where $Q_L^{1,2}$ and $Q_R^{1,2}$ are the charge $-1/3$ components of $Q_L^{1}$ and $Q_R^{1}$, respectively, while $Q_L^{2,1}$ and $Q_R^{2,1}$ are the charge $+5/3$ components of $Q_L^{2}$ and $Q_R^{2}$, respectively. In the charge $2/3$ sector, we have
\be
\mathcal L_{2/3}=-
\begin{pmatrix}
\bar S_L&\bar Q_L^{2,2}&\bar Q_L^{1,1}&\bar q_L
\end{pmatrix}
\begin{pmatrix}
\xi\hat fc_h&0&\mu_{tS}&0\\
\xi\hat f\frac{s_h}{\sqrt{2}}&\mu'_{QQ}&0&0\\
\xi\hat f\frac{s_h}{\sqrt{2}}&0&0&\mu_{QQ}\\
0&0&0&\mu_{qQ}
\end{pmatrix}
\begin{pmatrix}S_R\\Q_R^{2,2}\\t_R\\Q_R^{1,1}\end{pmatrix} + {\rm h.c.}~,
\ee
where $Q_L^{1,1}$, $Q_R^{1,1}$, $Q_L^{2,2}$ and $Q_R^{2,2}$ are the charge $2/3$ components of $Q_L^{1}$, $Q_R^{1}$, $Q_L^{2}$ and $Q_R^{2}$, respectively.
All phases can be reabsorbed into the different fields, so we take all parameters real and positive.
Only the charge $\frac{2}{3}$ fermions contribute to the effective potential. One can evaluate the characteristic polynomial of the matrix $\mathcal M_F$ appearing in $\mathcal L_{2/3}$
\be
\chi(p^2)\equiv \det \mathcal (\mathcal M_F^\dagger \mathcal M_F-p^2)=\chi_0(p^2)+s_h^2\, \chi_1(p^2)~,
\label{chi}
\ee
with
\bea
\chi_0&=&p^2(p^2-\xi^2\hat f^2-\mu_{tS}^2)(p^2-\mu_{QQ}'^2)(p^2-\mu_{QQ}^2-\mu_{qQ}^2) 
\nonumber\\
\chi_1&=& \frac{ \xi^2\hat f^2}{2}\, \biggl(
p^4[2\mu_{tS}^2-\mu_{QQ}^2-\mu_{QQ}'^2]\biggr.\nn\\&&\biggl.
+p^2[(2\mu_{QQ}^2+\mu_{qQ}^2)\mu_{QQ}'^2-(\mu_{QQ}^2+2\mu_{qQ}^2+\mu_{QQ}'^2)\mu_{tS}^2]+\mu_{tS}^2 \mu_{QQ}'^2\mu_{qQ}^2\biggr)~.
\eea
The first term corresponds to the unperturbed mass eigenvalues, the second term encodes the effects of EWSB. 
Notice that Eq.~(\ref{chi}) is exact as $\chi(p^2)$ does not have terms beyond quadratic order in $s_h$.

In terms of $\chi$ it is easy to write down the fermion contribution to the effective potential
\be
V_F=-\frac{N_c}{8\pi^2}\int d p^2\, p^2\, \log\, \chi(-p^2)\, .
\label{fermionpot}
\ee
One obtains 
\be
\alpha_{1/2}=\frac{3}{4\pi^2}\int dp^2 p^2\,  \frac{\chi_1(-p^2)}{\chi_0(-p^2)}\,,\qquad
\beta_{1/2}=\frac{3}{4\pi^2}\int dp^2 p^2 \left(\frac{\chi_1(-p^2)}{\chi_0(-p^2)}\right)^2\,,
\ee
One notices that $\alpha_{1/2}$ is IR finite but has a logarithmic UV divergence proportional to $\mu_{\rm eff}^2=2\mu_{tS}^2-\mu_{QQ}^2-\mu_{QQ}'^2$ (in the limit $\mu_{qQ}\to \infty$ the divergence is instead proportional to  $\tilde\mu_{\rm eff}^2=2\mu_{tS}^2-\mu_{QQ}'^2$), as argued diagrammatically in Section~\ref{GBpot}.
Conversely, $\beta_{1/2}$ is UV finite but logarithmically IR divergent, to be regularized by the top mass.

It is  useful to explicitly extract the UV divergent piece from $\alpha_{1/2}$ by writing 
$
\alpha_{1/2}=\alpha_{1/2}^{\rm div}+\alpha_{1/2}^{\rm fin}
$
with~\footnote{The choice of the IR scale $m_\H$ in this decomposition is done for later convenience.}
\be
\alpha_{1/2}^{\rm div}=\frac{3}{4\pi^2}\int dp^2 \,\frac{\xi^2\hat f^2}{2}\frac{\mu_{\rm eff}^2}{p^2+m_\H^2}
\,,\qquad
\alpha_{1/2}^{\rm fin}=\frac{3}{4\pi^2}\int dp^2 
\left(
p^2\,  \frac{\chi_1(-p^2)}{\chi_0(-p^2)}
-\frac{\xi^2\hat f^2}{2}\frac{\mu_{\rm eff}^2}{p^2+m_\H^2}
\right)
\ee
In the $\overline{MS}$ scheme one obtains 
\be
\alpha_{1/2}^{\rm div}=\frac{3}{8\pi^2}\xi^2\hat f^2\mu_{\rm eff}^2\left(
\log \frac{M^2}{m_\H^2}+1
\right)
\ee
where $M$ is the RG scale.

While the UV finite integrals in $\alpha^{\rm fin}_{1/2}$ and $\beta_{1/2}$ can be performed straightforwardly, the expressions are rather cumbersome so we do not report them here.

%
%

\subsection{Scalar Resonance}
\label{app:scalar}

The scalar sector also contributes to the NGB potential due to the $SO(5)_L$ violating coupling $\delta m^2$ introduced in Section~\ref{GBpot}.
The scalar mass matrix is given by
\be
\mathcal M_S^2~=~\begin{pmatrix}
(m_4^2+\lambda\H^2)\ 1_{3\times 3}\\
&m_4^2+\lambda(1+2s_h^2)\H^2&2\lambda s_hc_h\H^2\\
&2\lambda s_hc_h\H^2&m_1^2+\lambda(3-2s_h^2)\H^2
\end{pmatrix}~.
\ee
The vacuum expectation value of the radial field $\H$ is given by $\hat f^2=-m_1^2/\lambda$. In this vacuum, one obtains the effective potential
\be
V_S~=~\frac{1}{32\pi^2}\int dp^2 p^2\log\left([(p^2+\delta m^2+m_\H^2 s_h^2)(p^2+m_\H^2c_h^2)-m_\H^4s_h^2c_h^2]
\right)~,
\ee
and hence
\bea
\alpha_{0} &=&-\delta m^2 \hat{f}^2 +\frac{1}{16\pi^2}m_\H^2\delta m^2\left(\log\frac{M^2}{m_\H^2}+1\right)~,
\nn\\ [0.4em]
\beta_{0} &=&\frac{1}{16\pi^2}(\delta m^2)^2\left(1+\log\frac{m_\H^2(h)}{m_\H^2}\right)~,
\eea
where $m_\H$ is the Higgs mass that we used as an IR cutoff here. 
Notice that $\beta_{0}$ is of higher order in $\delta m^2$ and we will hence neglect it. 

\section{Expansion in Powers of $s_h$ and Logarithmic Divergences}
\label{app:IRRegulation}

In this appendix we clarify the issue of (spurious) IR divergences introduced when naively expanding the Coleman-Weinberg potential in powers of $x=s_h^2$. These arise from an unjustified expansion in the low-momentum region of integration, and is connected to states that acquire mass only as a result of $x \neq 0$. We use the spin-1/2 case for illustration, but the same considerations and conclusions hold true in the spin-1 and spin-0 sectors.

The 1-loop effective potential takes the form $V=-\frac{N_c}{8 \pi^2} \, \mathcal V$, where
\be
\mathcal V=\int_0^\infty \! d^2p \, p^2 \log \left(f(p^2)+g(p^2) \, x +\frac{1}{2} \, h(p^2) \, x^2 + \cdots \right)~,
\ee
and
\be
f(p^2)=p^2 f'_0+\cdots~,\quad g(p^2)=g_0+\cdots~, \quad h(p^2)=h_0+ \cdots~.
\ee
The expansion in small $x$  is then (neglecting a $x$-independent term):
\be
\mathcal V=x\ \lim_{x\to 0} \int_0^\infty \! d^2p \, p^2 \, \frac{g(p^2)}{f(p^2)} +\frac{x^2}{2} \ \lim_{x\to 0} \int_0^\infty \! d^2p \, p^2 \frac{f h-g^2}{(f+g x)^2}+ \cdots~.
\ee
The first term is IR safe so we can directly take the limit $x\to 0$.
In the second term we first make the shift of variables $p'^2=p^2+x g_0/f_0'$, and expand the integrand in $x$ to obtain
\be
\mathcal V=x \int_0^\infty \! d^2p \, p^2 \, \frac{g(p^2)}{f(p^2)} +
\frac{x^2}{2} \lim_{x\to 0} \int_{m^2_0}^\infty \! d^2p \, p^2 \left[\frac{f(p'^2) h(p'^2)-g^2(p'^2)}{f^2(p'^2)}+\mathcal O\left(\frac{x}{p'^4}\right)\right]+ \cdots~, 
\ee
where $m^2_0 = (g_0 / f'_0) \, s_h^2$ is the  (field-dependent) ``zero-mode'' mass squared.
The fist term in the brackets gives a contribution from the lower limit of integration of the form $ A \log(m^2_0)+B$.
The $\mathcal O\left(\frac{x}{p'^4}\right)$ term in the brackets gives only terms of order $x\log x$ and hence vanishes in this limit. The upshot is that one can obtain the correct small $x$ expansion by naively expanding the integrand, but cutting off the resulting IR divergence at the (field dependent) ``zero-mode'' mass, $m_0$. In practice, using the top (or $W$ or Higgs) mass at the minimum of the potential, i.e.~the physical mass, does not introduce a large difference compared to the more correct field-dependent one (see also \cite{Marzocca:2012zn}). 

\section{RG Equations in the Presence of QCD Interactions}
\label{app:RGeqs}

In this appendix we collect the RG equations used in the generation of Figs.~\ref{RGplot} and \ref{RGplot2}, meant to illustrate the effect of the gauge interactions on the fixed point behavior discussed in the main text:
\bea
\beta_{\xi^2}&=& \frac{\xi^2}{16\pi^2} \, \left\{ (17+N)\xi^2 -16 g_3^2 \right\}~,\nn\\
\beta_\lambda&=& \frac{1}{16\pi^2} \,  \left\{ 2(N+8)\lambda^2-24\xi^4 +24\xi^2\lambda \right\}~,\nn\\
\beta_{\muext^2} &=& \frac{\muext^2}{16\pi^2} \, \left\{ \xi^2 -16 g_3^2 \right\}~,\nn\\ 
\beta_{\delta m^2} &=& \frac{3\xi^2}{4\pi^2} \, \muext^2+\frac{\lambda+3\xi^2}{4\pi^2} \, \delta m^2~,\nn
\eea
where $g_3$ is the QCD coupling constant and we have taken $N_c = 3$.

In the main text we also used that
\be
\gamma_{L}~=~\frac{\xi^2}{32\pi^2}~,\qquad 
\gamma_{R}~=~\frac{N\xi^2}{32 \pi^2}~,\qquad 
\gamma_{\Phi}~=~\frac{N_c\xi^2}{8\pi^2}~,
\label{gammas}
\ee
where $\gamma_L$ is the anomalous dimension of $F_L$ and $\gamma_R$
that of $S_R$.  Using the Landau gauge ($\xi = 0$) for QCD, there are no 1-loop contributions to the fermion anomalous dimensions from $g_3$.

\bibliographystyle{JHEP}
\bibliography{biblioTop}

\end{document}